\documentclass[useAMS,usenatbib]{mn2e}
\usepackage{txfonts}
\usepackage{graphics}
\usepackage{times}

\def\del#1{{}}

\newcommand{\aj}{AJ}
\newcommand{\apj}{ApJ}
\newcommand{\apjs}{ApJ Suppl.}
\newcommand{\mnras}{MNRAS}

\newcommand{\aap}{A{\&}A}

\newcommand{\prd}{Phys. Rev. D}

\newcommand{\nat}{Nature}

\newcommand{\msun}{\rm\,M_\odot}
\newcommand{\kms}{\,{\rm km}{\rm s}^{-1}}
\newcommand{\hompc}{\,h\,{\rm Mpc}^{-1}}
\newcommand{\mpcoh}{\,h^{-1}\,{\rm Mpc}}

\newcommand{\simgt}%
{\,\hbox{\lower0.6ex\hbox{$\sim$}\llap{\raise0.6ex\hbox{$>$}}}\,}
\newcommand{\simlt}%
{\,\hbox{\lower0.6ex\hbox{$\sim$}\llap{\raise0.6ex\hbox{$<$}}}\,}

\begin{document}

\title[Redshift-space distortions]{Testing Cosmological Structure
  Formation using Redshift-Space Distortions}

\author[W.J. Percival and M. White]{
\parbox{\textwidth}{
Will J. Percival$^{1}$\thanks{E-mail: will.percival@port.ac.uk} and
Martin White$^{2}$}
\vspace*{4pt} \\
$^{1}$ Institute of Cosmology and Gravitation, University of
Portsmouth, Portsmouth, P01 2EG, UK \\
$^{2}$ Departments of Physics and Astronomy, University of California,
Berkeley, CA 94720, USA}

\date{\today} 
\pagerange{\pageref{firstpage}--\pageref{lastpage}}

\maketitle
\label{firstpage}

\begin{abstract}
  Observations of redshift-space distortions in spectroscopic galaxy
  surveys offer an attractive method for observing the build-up of
  cosmological structure. In this paper we develop and test a new
  statistic based on anisotropies in the measured galaxy power
  spectrum, which is independent of galaxy bias and matches the matter
  power spectrum shape on large scales. The amplitude provides a
  constraint on the derivative of the linear growth rate through
  $f\sigma_8({\rm mass})$. This demonstrates that spectroscopic galaxy
  surveys offer many of the same advantages as weak lensing surveys,
  in that they both use galaxies as test particles to probe all matter
  in the Universe. They are complementary as redshift-space
  distortions probe non-relativistic velocities and therefore the
  temporal metric perturbations, while weak lensing tests the sum of
  the temporal and spatial metric perturbations. The degree to which
  our estimator can be pushed into the non-linear regime is considered
  and we show that a simple Gaussian damping model, similar to that
  previously used to model the behaviour of the power spectrum on very
  small scales, can also model the quasi-linear behaviour of our
  estimator. This enhances the information that can be extracted from
  surveys for $\Lambda$CDM models.
\end{abstract}


\section{Introduction}

It is now widely accepted that the large-scale structure we see in the
distribution of galaxies arises through a process of gravitational
instability, which amplifies primordial fluctuations laid down in the
very early Universe. The rate at which structure grows from these
small perturbations offers a key discriminant between cosmological
models. For instance, dark energy models in which general relativity
is unmodified predict different Large-Scale Structure formation
compared with Modified Gravity models with the same background
expansion \citep[e.g.][]{dvali00,carroll04,Bra05,NesPer08}.

Structure growth is driven by the motion of matter (and inhibited by the
cosmological expansion). Galaxies are expected to act as test particles
within this matter flow, so the motion of galaxies carries an imprint
of the rate of growth of large-scale structure. Because of this, many
previous analyses have shown that observations of these galaxy peculiar
velocities can distinguish between classes of models
\citep[e.g.][]{jain07,song08,song08b}. A key technique to
statistically measure the growth of the velocity field uses
redshift-space distortions seen in galaxy surveys \citep{Kai87}. Galaxy
maps, produced by estimating distances from redshifts obtained in
spectroscopic galaxy surveys, reveal an anisotropic galaxy
distribution. The anisotropies arise because galaxy recession
velocities, from which distances are inferred, include components from
both the Hubble flow and peculiar velocities from the comoving motions
of galaxies. These distortions encode information about the build-up
of structure.

Many previous surveys have been analysed to measure
$\beta\approx\Omega_m^{0.6}/b$, where $b$ is the deterministic, local,
linear bias of the galaxies.  The latest generation of large surveys
have provided ever tighter constraints.
Analyses using the 2-degree Field Galaxy Redshift Survey
\citep[2dFGRS;][]{colless03} have measured redshift-space distortions
in both the correlation function \citep{peacock01,hawkins03} and power
spectrum \citep{percival04}.
Using the Sloan Digital Sky Survey \citep[SDSS;][]{york00}, redshift-space
distortions have also been measured in the correlation function
\citep{Zeh05,Oku08,cabre08}, and using an Eigenmode decomposition to separate
real and redshift-space effects \citep{tegmark04,tegmark06}.
These studies were recently extended to $z\simeq1$ \citep{guzzo08} using the
VIMOS-VLT Deep Survey \citep[VVDS;][]{lefevre05,garilli08} and the 2SLAQ
survey \citep{Ang08}.
In addition to measuring $\beta$ at $z=0.8$, \citet{lefevre05} emphasised the
importance of using large-scale peculiar velocities for constraining models
of cosmic acceleration.

On linear scales the theory behind the observed redshift-space
distortions is well developed \citep{Kai87,HamiltonReview},
and previous analyses have made use of this theory to measure $\beta$.
On quasi-linear and non-linear scales we are instead reduced to making
approximations, or using fitting formulae based on numerical simulations.
The standard approach is to use a `streaming' model, where linear
theory is spliced together with an approximation for random motion
of particles in collapsed objects (see section~\ref{sec:zspace_theory}).
\citet{TinWeiZhe06} and \citet{Tin07}, which are developments of work in
\citet{HatCol99}, discuss a number of possible improvements to the streaming
model based on fits to numerical simulation results.
These approaches aim to allow us to extract information on $\beta$ from
scales where the power spectrum does not match its linear form.
One concern is that the theoretical dependence on $\beta$ might change in
the quasi-linear regime, leading to complicated dependencies and that
fits to simulations will only be correct in a subset of cosmologies and
galaxy formation models similar to those from which the fits were
derived\footnote{The fits are quite sensitive to the satellite fraction,
for example.}.

A better approach would be to analyse the physics behind
redshift-space distortions, and to try to use this to devise the best
estimator with which to extract cosmological information. Because
galaxy velocities only depend on the distribution of matter, we can
devise an estimator $\hat{P}(k)$ that is not affected by galaxy bias:
instead it measures the large-scale shape of the matter power
spectrum. Because the estimator is based on the velocity power
spectrum, the large-scale amplitude is $f^2$ times that of matter
fluctuations, where $f$ is the logarithmic derivative of the linear
growth rate with respect to the scale factor, $f\equiv d\ln D/d\ln
a$. From this estimator, we can measure $f\sigma_{8,\,{\rm mass}}$,
which is proportional to $dD/d\ln a$, and provides a good discriminant
between modified gravity and dark energy models \citep{song08b}.
$\hat{P}(k)$ is expected to match the linear matter power spectrum
shape on large scales, so small-scale differences will help us to
determine the scales on which the theory is breaking down.  For
$\Lambda$CDM models, we show that a simple Gaussian smoothing of the
redshift-space power along the line-of-sight is able to match most of
this break-down. This places redshift-space distortion measurements on
the same footing as weak lensing measurements in the sense that they
both allow us to test the matter distribution directly. They provide
complementary information, as non-relativistic velocity measurements
only depend on temporal metric perturbations, while weak lensing tests
the sum of the temporal and spatial metric perturbations.

The layout of our paper is as follows. We first review the theory
behind redshift-space distortions (Section~\ref{sec:zspace_theory}) in
the linear, quasi-linear and non-linear regimes. In
Section~\ref{sec:estimators}, we introduce a new estimator, based on
the monopole and quadrupole from a Legendre decomposition of the
redshift-space power spectrum, which is designed to recover a power
spectrum given by $f^2$ times the matter power spectrum on large
scales. Section~\ref{sec:sims} introduces the N-body simulation used
to test this theory and our new estimator. We use spherically averaged
power spectra where we include distortions along multiple axes to
analyse this simulation, an approach described in
Section~\ref{sec:pk_sph_av}. Our analytic theory is compared with
results from the numerical simulation in
Section~\ref{sec:sim_results}. The paper ends with a discussion of our
results.

\section{modelling galaxy clustering in redshift-space} 
\label{sec:zspace_theory}

The redshift-space position of a galaxy differs from its real-space
position due to its peculiar velocity,
\begin{equation} \label{eq:sx}
  {\bf s} = {\bf x} - u_z({\bf x})\hat{\bf z}
  \qquad ,
\end{equation}
where $u_z({\bf x})$ is the line-of-sight component of the galaxy
velocity (assumed non-relativistic) in units of the Hubble velocity,
and we have taken the line-of-sight to be the $z$-axis.
We shall adopt the ``plane-parallel'' approximation, so this direction is
fixed for all galaxies.

The galaxy overdensity field in redshift-space can be obtained by
imposing mass conservation, $(1+\delta_g^s)d^3s=(1+\delta_g)d^3r$, and
the exact Jacobian for the real-space to redshift-space transformation
is
\begin{equation}
  \frac{d^3s}{d^3r} = \left(1+\frac{u_z}{z}\right)^2
    \left(1+\frac{du_z}{dz}\right)
  \qquad .
\end{equation}
In the limit where we are looking at scales much smaller than the mean distance
to the pair, $u_z/z$ is small and it is only the second term that is important
(\citealt{Kai87}; but see \citealt{PapSza08}),
\begin{equation}
  1+\delta_g^s = \left(1+\delta_g\right)\left(1+\frac{du_z}{dz}\right)^{-1}
  \qquad .
\end{equation}
If we assume an irrotational velocity field we can write
$u_z=\partial/\partial z\,\nabla^{-2}\theta$,
where $\theta\equiv\nabla\cdot{\bf u}$, and $\nabla^{-2}$ is the inverse
Laplacian operator. In Fourier space, $(\partial/\partial
z)^2\nabla^{-2}=(k_z/k)^2=\mu^2$, where $\mu$ is the cosine of the
line-of-sight angle, so we have that
\begin{equation}  \label{eq:del_full}
  \delta_g^s(k) = \delta_g(k) - \mu^2\theta(k) 
    - \mu^2[\delta\otimes\theta](k) 
    + \mu^4[\theta\otimes\theta](k) + \cdots
  \qquad ,
\end{equation}
including second order convolutions in $\theta(k)$ and $\delta_g(k)$,
while neglecting third and higher order terms.

\subsection{The linear regime}  \label{sec:lin}

\begin{figure}
\centering
\resizebox{0.9\columnwidth}{!}{\includegraphics{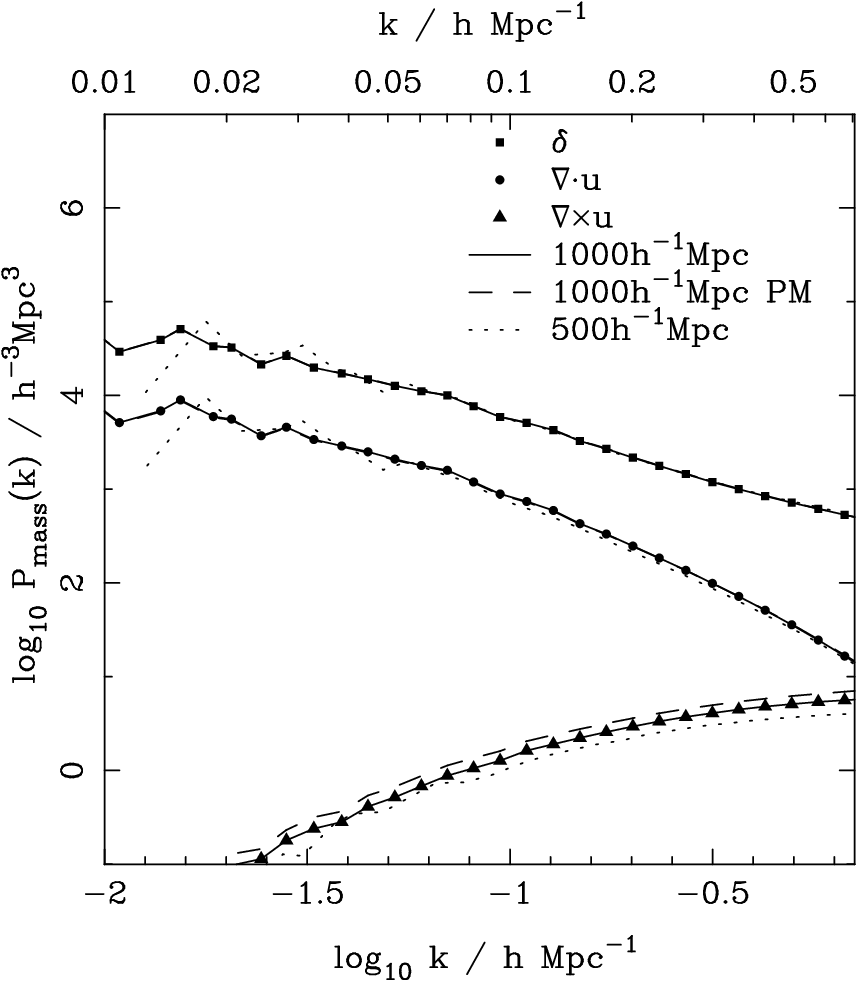}}
\caption{Mass density and velocity power spectra recovered from three
  different simulations. Solid lines correspond to the primary
  simulation used in this paper, and for this simulation we show power
  spectra of the overdensity (solid squares), velocity divergence
  (solid circles) and the curl of the velocity (solid triangles). For
  comparison, the dashed line is the corresponding power spectra from
  a simulation with the same initial conditions and cosmology, run
  with a PM code (so lower force resolution), and the dotted lines
  corresponds to a simulation with a box half as big (so twice the
  force resolution, 8 times the mass resolution but 1/8 the volume).}
\label{fig:pksim}
\end{figure}

If $\theta$ and $\delta_g$ are small, then we can drop the second and
higher order terms from Eq.~(\ref{eq:del_full}), and
\begin{equation}  \label{eq:del_mu}
  \delta_g^s(k) = \delta_g(k) - \mu^2 \theta(k)
  \qquad .
\end{equation}

Often it is further assumed that the velocity field comes from linear
perturbation theory. Then
\begin{equation}  \label{eq:lin_f}
  \theta(k) = -f \delta_{\rm mass}(k)
  \qquad ,
\end{equation}
where $f\equiv d\ln D/d\ln a \approx \Omega_m^{0.6}$ \citep{Pee80}.

For a population of galaxies, which we denote with a subscript $g$,
the linear redshift-space power spectrum can be written
\begin{eqnarray} 
  P_g^s(k,\mu,{\rm lin}) &\equiv& 
    \langle |\delta_g^s(k,\mu)|^2 \rangle,\\
  &=& P_{gg}(k)-2\mu^2P_{g\theta}(k)+\mu^4P_{\theta\theta}(k)
  \qquad ,  \label{eq:pgs_lin}
\end{eqnarray}
where $P_{gg}(k)\equiv\langle |\delta_g({\bf k})|^2\rangle$,
$P_{g\theta}(k)\equiv\langle\delta_g({\bf k})\theta({\bf k})\rangle$,
$P_{\theta\theta}(k)\equiv\langle |\theta({\bf k})|^2\rangle$, are the
galaxy--galaxy, galaxy--$\theta$ and $\theta$--$\theta$ power spectra
respectively for modes ${\bf k}$. A subscript $\theta$ shows that a
variable is determined from the velocity field of the galaxies, which
we have assumed is irrotational and small compared with the real-space
distance to the galaxies (see Fig.~\ref{fig:pksim}). In the following
we often drop explicitly showing the $k$ dependence of these power
spectra, for convenience.

\subsection{Galaxy bias}

It has long been known that galaxies do not trace the mass, a phenomenon
known as `bias'.  There are good theoretical reasons to believe that on
large scales the shape of the power spectra of galaxies and mass are
similar \citep[e.g.][]{Pee80,Pea99}, which can be phrased as the assumption of
scale-independent bias.
Under this approximation, \citet{dekel99} introduced the bias relation,
\begin{equation} \label{eq:bias_dekel}
    \left( \begin{array}{cc}
        \langle\delta({\bf x})\delta({\bf x})\rangle &
        \langle\delta({\bf x})\delta_g({\bf x})\rangle \\
        \langle\delta_g({\bf x})\delta({\bf x})\rangle &
        \langle\delta_g({\bf x})\delta_g({\bf x})\rangle
    \end{array}\right)
  = 
    \langle |\delta_{\rm mass}({\bf x})|^2 \rangle
    \left( \begin{array}{cc} 1 & br \\ br & b^2 \end{array}\right),
\end{equation} 
where $b$ is the bias factor, and $r$ is the dimensionless correlation
coefficient between the distributions of mass and galaxies. If $r=1$,
we have a fully deterministic local linear bias relation,
\begin{equation}  \label{eq:lin_bias}
  \delta_g({\bf x}) = b \delta_{\rm mass}({\bf x})
  \qquad .
\end{equation}
If biasing is not a purely Poisson process, but has a random
stochastic element, then we could have $r<1$. Measurements of $r$,
such as summarized in Figure 11 of \citet{swanson08}, indicate that $r$
is close to 1 on large scales.

If $b\simeq1$, or $\delta({\bf x})$ and $b\delta({\bf x})$ are small,
then galaxy velocities randomly sample those of the matter in the
Universe and we can write,
\begin{eqnarray}  
  P_{\theta\theta}(k) &=& f^2P_{\rm mass}(k) \nonumber \\
  P_{g\theta}(k)     &=& -bfrP_{\rm mass}(k), \label{eq:Pmass_relations}
\end{eqnarray}
where $P_{\rm mass}(k)\equiv\langle|\delta(k)|^2\rangle$.  This is not
true for some models of galaxy bias, such as the peaks model, which
predicts a velocity-bias where the distribution of galaxy velocities
does not match that of the mass \citep{regos95,percival08}.  Under the
assumptions that lead to Eqns.~\ref{eq:Pmass_relations}, the standard
way of writing the redshift-space galaxy power spectrum is
\begin{equation}  \label{eq:pgs_beta}
  P_g^s(k,\mu) = P_{gg}(k)\left[1+2\mu^2\beta+\mu^4\beta^2\right]
  \qquad ,
\end{equation}
where $\beta\equiv f/b$. By writing the power spectrum in this form we
are hiding a key feature of linear redshift-space distortions: that
they depend on the mass overdensity not the galaxy overdensity. To see
this, we now consider the normalisation of the power spectrum.

\subsection{Power spectrum normalisation}

The normalisation of power spectra is usually measured using the rms
fluctuation amplitude, in spheres of radius $8\mpcoh$, which would have
arisen if linear evolution had proceeded to the time of observation.
For the anisotropic redshift-space power spectrum we can define a slightly
non-standard, anisotropic normalisation
\begin{equation}
  \sigma_{8,gal}^2(\mu)\equiv \int \frac{dk}{2\pi^2} W_8^2(k)k^2P_g^s(k,\mu)
  \qquad ,
\end{equation}
where $W_8(k)$ is the Fourier transform of the top-hat window function
of width $8\hompc$. Here we have decomposed the standard 3D integral
into a 1D integral assuming statistical isotropy of the density field,
and have performed this integral along particular directions to the
line-of-sight. For each $\mu$, this is equivalent to the standard
definition of $\sigma_8$ for an isotropic power with the same
amplitude as $P_g^s(k,\mu)$. Substituting Eq.~(\ref{eq:pgs_lin}) into
this expression, and using the fact that $P_{gg}$, $P_{g\theta}$ \&
$P_{\theta\theta}$ are isotropic, we have that
\begin{equation}  \label{eq:sig8}
  \sigma_{8,\,g}^2(\mu) = (b\sigma_{8,\,{\rm mass}})^2 +
    2\mu^2(br\sigma_{8,\,{\rm mass}})(f\sigma_{8,\,{\rm mass}}) +
    \mu^4(f\sigma_{8,\,{\rm mass}})^2
  \quad .
\end{equation}
Because of the $\mu$ dependence, large-scale peculiar velocity
observations provide a measurement of $f\sigma_{8,\,{\rm mass}}$ that
is independent of a local linear bias.  From observations of the
large-scale amplitude of $P_g^s(k,\mu)$ as a function of $\mu$, we can
hope to measure $f\sigma_{8,\,{\rm mass}}$, $b\sigma_{8,\,{\rm
    mass}}$, and $br\sigma_{8,\,{\rm mass}}$.  Eq.~(\ref{eq:sig8})
shows in compact form that we cannot break these parameter
combinations and constrain $f$, $b$ or $\sigma_{8,\,{\rm mass}}$
independently solely using $P_g^s(k,\mu)$, although we can measure
$r$.

\subsection{The quasi-linear regime}  \label{sec:qlin}

In addition to removing higher order terms, the two assumptions
leading to Eq.~(\ref{eq:del_full}) are that we are dealing with an
irrotational velocity field, and that we are in the distant observer
limit ($u_z\ll z$) both of which we assume to remain true in the
quasi-linear regime. Such assumptions (among others) are also made in
standard Eulerian perturbation theory
\citep{Jus81,Vis83,Fry84,Gor86,Mak92,JaiBer94,Ber02},
Lagrangian perturbation theory \citep{Buc89,Hiv95},
renormalised perturbation theory \citep{Sco01}
and resummed Lagrangian perturbation theory \citep{Mat07}.

Going beyond the linear assumption leads to additional terms in
$\mu^2$, $\mu^4$ \& $\mu^6$ in the redshift-space power spectrum
compared with Eq.~(\ref{eq:pgs_lin}). So,
\begin{equation}  \label{eq:pgs_qlin}
  P_g^s(k,\mu,{\rm qlin}) = \sum_{i=0}^{3}A_i(k)\mu^{2i}
  \qquad ,
\end{equation}
where $A_1\ne P_{g\theta}$, $A_2\ne P_{\theta\theta}$, and $A_3\ne0$.
The importance of the extra higher order terms was emphasised most
recently by \citet{Sco04}.  In addition to these redshift-space
effects, the linear theory relation between $\delta$ and $\theta$,
given in Eq.~(\ref{eq:lin_f}), will break down in the quasi-linear
regime, so we should expect the shapes of $P_{gg}$, $P_{g\theta}$ and
$P_{\theta\theta}$ to be different.  The relationship between
Eqns.~(\ref{eq:pgs_lin} \&~\ref{eq:pgs_qlin}) can be written
\begin{equation}  \label{eq:G}
  P_g^s(k,\mu,{\rm qlin}) = G(k,\mu^2) P_g^s(k,\mu,{\rm lin})
  \qquad ,
\end{equation}
where $G(k,\mu^2)$ has the property $\lim_{k\to0}G(k,\mu^2)=1$. 

\subsection{The non-linear regime} \label{sec:nlin}

The standard model for redshift-space distortions includes a component
caused by an uncorrelated velocity dispersion that grows on small
scales. Such a model is motivated by the idea of ``thermal motion'' of
particles in collapsed structures, which causes the Fingers-Of-God
(FOG) observed in redshift surveys \citep{Jac72}.  An additional
component comprising uncorrelated particle motions will dilute both
the galaxy overdensity $\delta_g$, and the ``extra'' overdensity term
caused by the linear distortions, $\theta$.  Motivated by numerical
simulations \citep{SheDia01,HSWSW07} and the halo model
\citep{Whi01,Sel01}, we can assume that the centre of mass of a halo,
around which galaxies orbit, still moves according to (quasi-)linear
motion.  Such a model leads to the much-used `streaming' models
\citep[e.g.][]{HamiltonReview} where
\begin{equation}
  P_g^s(k,\mu,{\rm nl}) = F(k,\mu^2) P_g^s(k,\mu,{\rm lin})
  \qquad ,
\end{equation}
with $F(k,\mu^2)$ a function that depends on the distribution of
random pair velocities in collapsed objects, which is often written as
a function of $y=k\sigma$, where $\sigma$ is the rms velocity
dispersion. In order to match behaviour on large scales, we require
$\lim_{k\to0}F(k,\mu^2)=1$. We have written the equation in this form
to highlight the similarity with Eq.~(\ref{eq:G}).
Note that this model is constructed by a rather ad-hoc splicing of linear,
quasi-linear and non-linear behaviour which ignores the scale-dependence
of the mapping between real and redshift space separations
\citep[][and references therein]{Fis95,Sco04},
while Eq.~(\ref{eq:G}) was based on the analysis of the redshift-space
distortions in the quasi-linear limit.

In general FOG are difficult to model well, and their amplitude is strongly
dependent on the mean halo mass and satellite fraction of the population
under consideration \citep{Whi01,Sel01}.
Previous work has concentrated on models with Gaussian or Exponential
distributions \citep[e.g.][]{cole95,PeaDod96} for the pairwise
velocity dispersion in configuration space. For an Exponential model
for the pairwise velocity dispersion in configuration space, we expect
a Lorentz damping factor for the power spectrum, while the Gaussian
dispersion translates to a Gaussian damping of the power spectrum
\begin{eqnarray}
  F_{\rm Exponential}(k,\mu^2) &=& \left[1+(k\sigma\mu)^2\right]^{-1} 
    \label{eq:F_exp} \qquad , \\
  F_{\rm Gaussian}(k,\mu^2) &=& \exp\left[-(k\sigma\mu)^2\right]
    \label{eq:F_gauss} \qquad .
\end{eqnarray}
These terms have the same behaviour to first order. 

The exact form of $F(k,\mu^2)$, and the value of $\sigma$ is strongly
dependent on the galaxy population \citep{Jing04,Li07}. An alternative
approach would be to try to ``eliminate'' the FOG by applying a halo
finding algorithm to the sample and manually moving galaxies either to
halo centres, or to a spherically symmetric distribution around these
centres (e.g. \citealt{tegmark04}). Such approaches tend to mask the
fact that the radial ``compression'' is still model dependent, and
requires a similar free parameter to $\sigma$ in
Eq.~(\ref{eq:pgs_combined}). This ``parameter'' controls the
probability density function for the distortion of any galaxy in
redshift space \citep{reid08}. However, FOG compression does have the
advantage of including extra information in the analysis from the
phases, which are used to locate the halos.

Dealing directly with the FOG is not the same as extending the linear
model, and $P_{gg}$, $P_{g\theta}$ \& $P_{\theta\theta}$ into the
non-linear regime, as discussed in the previous section.  The
real-space effect of random thermal motion of galaxies on small scales
would lead $P_{g\theta}$ to decrease in amplitude, because of the
decoherence of density and velocity divergence, while
$P_{\theta\theta}$ increases.  We showed in Eq.~(\ref{eq:G}) that a
similar function to $F(k,\mu^2)$ would be required to include
quasi-linear behaviour in the redshift-space power spectrum. In this
case we would not expect that these simple damping models can
simultaneously match both the quasi-linear and fully non-linear
behaviour \citep{Fis95,Sco04}.

Clearly, if $F(k,\mu^2)$ has two roles, we must reconsider any implied
or assumed physical meaning in the function. So, for example, the
physical arguments that the Exponential form should be a better match
to the average velocity dispersion in halos
\citep{She96,PeaDod96,Whi01} are not applicable. For consistency with
the standard `streaming model', we continue to refer to the function
$F(k,\mu^2)$ as a small-scale velocity dispersion (SSVD) model.

\subsection{Combined model}

Following our review of the theory behind redshift-space distortions
presented in Sections~\ref{sec:lin}--\ref{sec:nlin}, we are left with
a model of the redshift-space galaxy power spectrum
\begin{equation} \label{eq:pgs_combined}
  P_g^s(k,\mu) = \left[P_{gg}(k) + 2 \mu^2 P_{g\theta}(k)
    + \mu^4 P_{\theta\theta}(k)\right] F(k,\mu^2).
\end{equation}
To first order in $k$, both the Gaussian and Exponential forms for
$F(k,\mu^2)$ considered in Section~\ref{sec:nlin} act as an extra
$\mu^6$ component with amplitude determined by $\sigma$. This fitting
function can act as an approximate correction for both quasi-linear
and non-linear terms in the mapping between linear real-space and
observed redshift-space. $P_{\theta\theta}$ is independent of galaxy
density bias, and is directly related to the matter velocity power
spectrum, provided there is no velocity bias, caused by a mis-match
between the distribution of galaxy velocities and the distribution of
velocities in all matter.

Eq.~(\ref{eq:pgs_combined}) could be fitted to data directly using a
Likelihood approach \citep{hamilton00,tegmark04,tegmark06}, resulting
in observational constraints on $P_{gg}$, $P_{g\theta}$, and
$P_{\theta\theta}$. Linking $P_{gg}$, $P_{g\theta}$, and
$P_{\theta\theta}$ to the matter power spectrum requires further
assumptions. Following the deterministic, linear local galaxy bias
model presented in Eq.~(\ref{eq:lin_bias}), and the assumption of no
velocity bias, the three power spectra are related to the matter power
spectrum according to Eqns.~\ref{eq:Pmass_relations}, which leads to a
model
\begin{equation} \label{eq:pgs_comb_lin}
  P_g^s(k,\mu) = [b^2 + 2 \mu^2 bf + \mu^4 f^2]P_{\rm mass}(k) F(k,\mu^2).
\end{equation}
This allows observations to constrain the matter power spectrum
multiplied by $f^2$. We show that this model matches simulation
results on large-scales in Section~\ref{sec:sim_results}. If
Eq.~(\ref{eq:pgs_comb_lin}) holds, Likelihood fits could be used to
measure $f\sigma_8({\rm mass})$ and $b\sigma_8({\rm mass})$ from the
measured power spectra. Alternatively, we show in the next Section
that estimators of the matter power spectrum multiplied by $f^2$ can
be constructed, based on a Legendre polynomial decomposition of the
redshift-space power spectrum.

\section{revised estimators for cosmological information} 
\label{sec:estimators}

We argued in Section~\ref{sec:lin} that large-scale linear
redshift-space distortions should enable us to measure $f\sigma_8({\rm
  mass})$ independently of linear bias. We will now show how we can do
so, based on a Legendre decompositon of the redshift-space power
spectrum. $P_g^s(k,\mu)$ can be decomposed into Legendre polynomials
$L_\ell(\mu)$ to give multipole moments
\begin{equation}  \label{eq:Pl_def}
  P_\ell^s(k)\equiv\frac{2\ell+1}{2}\int^{+1}_{-1}d\mu\ P_g^s(k,\mu)
    L_\ell(\mu)
  \qquad .
\end{equation}
The first three even Legendre polynomials are $L_0(\mu)=1$,
$L_2(\mu)=(3\mu^2-1)/2$ and $L_4(\mu)=(35\mu^4-30\mu^2+3)/8$. 
For redshift-space distortions along one axis, the expansion of the
linear model for $P_g^s(k,\mu)$ (Eq.~\ref{eq:pgs_lin}) into the first
three multipole moments gives
\begin{equation} \label{eq:pk_moments}
  \left(\begin{array}{c} 
    P_0^s(k) \\
    P_2^s(k) \\
    P_4^s(k) \end{array}\right) =
  \left(\begin{array}{ccc}
      1 &  2/3 & 1/5 \\
      0 & 4/3 & 4/7 \\
      0 & 0 & 8/35 \end{array}\right)
  \left(\begin{array}{c} 
    P_{gg}(k) \\
    P_{g\theta}(k) \\
    P_{\theta\theta}(k) \end{array}\right)
  \qquad ,
\end{equation}
so we see that the monopole, quadrupole and hexadecapole completely
characterise $P_g^s(k)$ in the linear limit.

If we assume that $P_{gg}$, $P_{g\theta}$ and $P_{\theta\theta}$ have
different forms, then we can invert Eq.~(\ref{eq:pk_moments}) to
recover them individually from $P_0^s$ $P_2^s$, and $P_4^s$,
\begin{equation} \label{eq:pk_from_legendre}
  \left(\begin{array}{c} 
    P_{gg}(k) \\
    P_{g\theta}(k) \\
    P_{\theta\theta}(k) \end{array}\right) =
  \left(\begin{array}{ccc}
      1 &  -1/2 & 3/8 \\
      0 & 3/4 & -15/8 \\
      0 & 0 & 35/8 \end{array}\right)
  \left(\begin{array}{c} 
    P_0^s(k) \\
    P_2^s(k) \\
    P_4^s(k) \end{array}\right)
  \qquad .
\end{equation}
In terms of $\mu$, $P_{gg}$, $P_{g\theta}$, and $P_{\theta\theta}$ can
be considered as revised moments by substituting the expressions for
$P_0^s$, $P_2^s$, \& $P_4^s$ into this expression.

The inclusion of a SSVD model given by Eq.~(\ref{eq:F_exp}) or
Eq.~(\ref{eq:F_gauss}) in the analysis of this statistic is considered
in \citet{cole95}: the integrals in Eq.~(\ref{eq:Pl_def}) need to be
recalculated as in Section~\ref{sec:sph_av_nl}. This would enable us
to use the monopole, quadrupole and hexadecapole to extract the same
information from simulations as using the spherically averaged power
spectra described in Section~\ref{sec:pk_sph_av}.

Following Eq.~(\ref{eq:pgs_beta}), measurements of $P_2^s/P_0^s$
have been previously used to measure $\beta$ through \citep{cole94}
\begin{equation}  \label{eq:P2overP0}
  \frac{P_2^s(k)}{P_0^s(k)}=\frac{ \frac{4}{3}\beta + \frac{4}{7}\beta^2}
    {1+\frac{2}{3}\beta+\frac{1}{5}\beta^2}
  \qquad .
\end{equation}
This only involves the ``lowest order'' moments in $\mu$ ($P_0^s$ and
$P_2^s$) and it has been argued that this makes it less sensitive to
noise. Expanding in multipoles can also simplify the definition of
simple survey limits. The expansion of this formulae to include a SSVD
model has also been considered by \citet{cole95}. The quadrupole to
monopole ratio is expected to have this limiting behaviour on
large-scales, and can therefore be used to measure $\beta$.

We now show that we can use $P_0^s$ and $P_2^s$ to derive an estimator of
the matter power spectrum multiplied by $f^2$, which leads to
cosmological constraints that are independent of a local galaxy bias.
For a local linear bias, in the absence of velocity bias,
$P_{\theta\theta}$ and $(P_{g\theta})^2/P_{gg}$ both have
normalisation $f^2\sigma_8^2({\rm mass})$ and shape matching that of
the matter density power spectrum in the linear limit. Results from
simulations, presented in Section~\ref{sec:sim_results} are consistent
with this claim. W can therefore define a new estimator $\hat{P}$,
such that $\hat{P}=(P_{g\theta})^2/P_{gg}$, and
$\hat{P}=P_{\theta\theta}$ on large scales.

In terms of the Legendre decomposition, we can write two functions of
$P_0^s$, $P_2^s$ and $P_4^s$, which should have the same large-scale
normalisation that is independent of a local linear bias. In the
linear regime, from Eq.~(\ref{eq:pk_from_legendre}) these functions
are
\begin{eqnarray}
  \hat{P}(k) &=&
    \frac{[\frac{3}{4}P_2^s - \frac{15}{8} P_4^s]^2}
         {P_0^s-\frac{1}{2}P_2^s+\frac{3}{8}P_4^s} \\
  \hat{P}(k) &=& \frac{35}{8} P_4^s
  \qquad .
\end{eqnarray}
Because we have two equations for $\hat{P}$, we can eliminate $P_4^s$,
leaving a quadratic equation in $\hat{P}$
\begin{equation}
  24[\hat{P}(k)]^2-35(7P_0^s+P_2^s)\hat{P}(k)+\frac{2205}{16}(P_2^s)^2=0
  \qquad .
\end{equation}
Solving this equation gives
\begin{equation}  \label{eq:bsig8}
  \hat{P}(k) = \frac{7}{48}
    \left[5(7P_0^s+P_2^s)-\sqrt{35}[35(P_0^s)^2+10P_0^sP_2^s-7(P_2^s)^2]^{1/2}\right]
  \qquad ,
\end{equation}
which offers a mechanism for removing a local bias dependence from
quadrupole and monopole measurements in the distant observer
limit. The result is a power spectrum whose large-scale shape should
match that of the mass, and normalisation should be $f^2$ times that
of the mass density power spectrum.

We can extend this model to the quasi-linear regime by including the
model for small-scale velocity dispersion. As we argued previously, in
the quasi-linear regime, this should be considered to be a fitting
formula for quasi-linear distortions, and is not physically
motivated. In the following we only consider a Gaussian SSVD model,
because this is favoured by simulation results presented in the next
section. A similar formula could be calculated for the Exponential
SSVD model. In the linear--Gaussian model, we can simplify the
equations by defining
$\Gamma_n(y)\equiv\gamma([n+1]/2,y^2)/(2y^{n+1})$ so that
\begin{equation} \label{eq:pk_leg_gauss}
  \left(\begin{array}{c} 
    P_0^s \\
    P_2^s \end{array}\right) =
  \left(\begin{array}{ccc}
    \Gamma_0 & 2\Gamma_2 & \Gamma_4 \\
    \frac{5}{2}(3\Gamma_2-\Gamma_0) & 
        15\Gamma_4-5\Gamma_2 & 
        \frac{5}{2}(3\Gamma_6-\Gamma_4) \end{array}\right)
    \left(\begin{array}{c} 
        P_{gg} \\
        P_\times \\
        \hat{P} \end{array}\right)
\end{equation}
where we have defined $P_\times\equiv\sqrt{P_{gg}\hat{P}}$ for
notational convenience.  This equation reduces to the relevant
expression in Eq.~(\ref{eq:bsig8}) in the limit as $y\to0$.  As
outlined above for the linear case, these two equations can be
manipulated to eliminate $P_{gg}$ and calculate an expression for
$\hat{P}$, which is a quadratic in $P_0^s$ and $P_2^s$, as in
Eq.~(\ref{eq:bsig8}) and can be easily solved, although it now depends
on $\Gamma_n(y)$ in a complicated way. The linear and damped versions
of this estimator are compared with the results from our numerical
simulation in Section~\ref{sec:sim_test_estimator}.

\section{Simulation}  \label{sec:sims}

To investigate redshift-space clustering further we have used a large,
high-resolution N-body simulation which is well suited to probing
$P_g^s(k,\mu)$ in the quasi-linear regime.  The cosmology was of the
$\Lambda$CDM family with $\Omega_m=0.25=1-\Omega_\Lambda$,
$\Omega_b=0.043$, $h=0.72$, $n_s=0.97$ and $\sigma_8=0.8$.  The linear
theory power spectrum for the initial conditions was computed by
evolution of the coupled Einstein, fluid and Boltzmann equations using
the code described in \citet{WhiSco95}.  \citet{SSWZ} find that this
code agrees well with CMBfast \citep{CMBfast}.  The simulation
employed $1024^3$ particles of mass $10^{11}\,h^{-1}~M_\odot$ in a
periodic cube of side $1\,h^{-1}$Gpc using a {\sl TreePM\/} code
\citep{Whi02} with a Plummer-equivalent softening length of
$35\,h^{-1}~$kpc (comoving).  A detailed comparison of this {\sl TreePM\/}
code with other codes can be found in \citet{Hei08} and \citet{Evr08}.

In order to test the convergence of this simulation, we compare
present-day density and velocity power spectra against alternative
simulations in Fig.~\ref{fig:pksim}.  We compare with a simulation run
from the same initial conditions, but using a particle-mesh code which
has lower force resolution.  We also compare against a simulation with
half the box size, i.e.~double the force resolution over $1/8$ of the
volume.  The density power spectra recovered from the simulations do
not reveal any significant trends with force resolution.  There is
weak evidence for smaller $\nabla\times{\bf u}$ and velocity
divergence with higher force resolution, but this is not significant
for our purposes.

The offset between density and velocity divergence power spectra seen
in Fig.~\ref{fig:pksim} on large scales is caused by the factor $f^2$
as described in Section~\ref{sec:lin}. The difference on smaller
scales is analysed in detail later. Our assumption that the velocity
field is curl-free is validated by these simulations, which show that
$\left|\nabla\times{\bf u}\right|_k \ll \left|\nabla\cdot{\bf u}\right|_k$
for the scales of interest: $k<0.2\hompc$.

From the $z=0$ output we generate a halo catalog using the
Friends-of-Friends (FoF) algorithm \citep{DEFW} with a linking length
of 0.168 times the mean inter-particle spacing.  This procedure
partitions the particles into equivalence classes, by linking together
all particle pairs separated by less than a distance $b$.  The halos
correspond roughly to particles with $\rho>3/(2\pi b^3)\simeq 100$
times the background density.  We will present results calculated from
all particles and from catalogues containing all particles in halos
more massive than $\log_{10} M=12.5$, 13 and 13.5 where $M$ is in
units of $h^{-1}M_\odot$. The small-scale velocity dispersions of the
particles will match the distribution of mass in each halo. We do not
include further small-scale effects caused by the galaxies inside each
halo not Poisson sampling the mass. Our intention in this paper is not
to try to model the non-linear regime perfectly, but to test the
effect of halo mass selection on the linear and quasi-linear
regimes. We therefore adopt this simple halo occupation distribution
in order to reduce the shot noise: by using all particles in each halo
we optimise our determination of the (quasi-)linear halo velocity.

We cannot easily replicate Fig.~\ref{fig:pksim} for our halo-particle
catalogues because of the difficulty of interpolating the velocity
field between the sparsely sampled galaxies. Instead, in the rest of
this paper, we directly analyse redshift-space distortions through
$P_g^s(k,\mu)$. We will consider two ways of analysing $P_g^s(k,\mu)$,
either using spherically averaged power spectra with different
redshift-space distortion components, or by decomposing $P_g^s(k,\mu)$
into Legendre polynomials. In the simulations we determine $P_\ell^s$
in each $k$-bin both by directly integrating against $\mu$ as in
Eq.~(\ref{eq:Pl_def}) and by a least-squares fit of $P_g^s(k,\mu)$ to
a sum of Legendre polynomials. The results are consistent. Before
presenting the results from simulations, we first consider the theory
behind decomposing simulation results into spherically averaged power
spectra.

\section{spherically averaged power spectra from simulations}  
\label{sec:pk_sph_av}

In this section we introduce the concept of spherically averaged power
spectra calculated from simulations where we propagate redshift-space
distortions along multiple axes. Clearly, such power spectra will not
match observational results, where redshift-space distortions are only
radial. However, they do provide a convenient and simple way to
explore the components of Eq.~(\ref{eq:pgs_combined}) using
simulations. For a sample of galaxies where we have applied
redshift-space distortions along multiple axes, for each ${\bf k}$,
Eq.~(\ref{eq:pgs_combined}) will still hold for some value of
$\mu$. The spherically averaged power spectra act as $\mu$-dependent
moments of $P_g^s(k,\mu)$, and this method does not alter the physics
that we are testing. We are simply summing these modes over different
combinations of $\mu$. When analysing simulations, we apply the
plane-parallel decomposition to both the simulation data and the
analysis method. Consequently, this should not be considered an {\em
  approximation} as we do not relate the results to actual surveys,
but only use simulations to investigate the physics.

\subsection{Linear distortions}

For redshift-space distortions along one axis, in the linear regime
where Eq.~(\ref{eq:del_mu}) holds, we have
\begin{equation} \label{eq:pklin_1axis}
  P_{1\,{\rm axis}}^s(k)
    =\int_0^1d\mu\ \left\langle[\delta_g(k)+\mu^2\theta(k)]^2\right\rangle
  \qquad .
\end{equation}
For two axes with redshift-space distortions, we can take $\nu$ to be
the cosine of the angle to the direction where there are no
distortions and integrate so that, in the linear limit,
\begin{equation}  \label{eq:pklin_2axes}
  P_{2\,{\rm axes}}^s(k)
    =\int_0^1d\nu\ 
    \left\langle[\delta_g(k)+\theta(k)-\nu^2\theta(k)]^2\right\rangle
  \quad .
\end{equation}
For three axes, whatever direction we take our $k$-vector, we see
redshift-space distortion of the overdensity is as in
Eq.~(\ref{eq:del_mu}) with $\mu=1$. Solving the integrals in
Eqns.(\ref{eq:pklin_1axis}) \&~(\ref{eq:pklin_2axes}), and
substituting in the relations $P_{gg}(k)\equiv\langle
|\delta_g(k)|^2\rangle$,
$P_{g\theta}(k)\equiv\langle\delta_g(k)\theta(k)\rangle$
\&~$P_{\theta\theta}(k)\equiv\langle|\theta(k)|^2\rangle$, leads to
the following spherically averaged power spectra,
\begin{equation} \label{eq:pk_sph_lin}
  \left(\begin{array}{c} 
    P_{0\,{\rm axis}}^s(k) \\
    P_{1\,{\rm axis}}^s(k) \\
    P_{2\,{\rm axes}}^s(k) \\
    P_{3\,{\rm axes}}^s(k) \end{array}\right) = \frac{1}{15}
  \left(\begin{array}{ccc}
      15 &  0 & 0 \\
      15 & 10 & 3 \\
      15 & 20 & 8 \\
      15 & 30 & 15 \end{array}\right)
  \left(\begin{array}{c} 
    P_{gg}(k) \\
    P_{g\theta}(k) \\
    P_{\theta\theta}(k) \end{array}\right)
  \quad .
\end{equation}
Here $P_{i\,{\rm axes}}^s(k)$ includes redshift-space distortions
along $i$ axes.  Using any three of $P_{i\,{\rm axes}}^s(k)$, we can
reconstruct $P_{gg}$, $P_{g\theta}$, and $P_{\theta\theta}$ by solving
the corresponding linear equations.

\subsection{Including small-scale velocity dispersion}  \label{sec:sph_av_nl}

If we include a SSVD damping model then, for a power spectrum with
redshift-space distortions along one axis, 
\begin{equation}  \label{eq:pk_one}
  P_{1\,{\rm axis}}^s(k)=\int_0^1d\mu
    \ \left\langle[\delta_g(k)+\mu^2\theta(k)]^2\right\rangle
    F(k,\mu^2)
  \quad .
\end{equation}
For redshift-space distortions along two axes, we take $\nu$ to be the
cosine of the angle to the direction where there are no distortions,
as in Eq.~(\ref{eq:pklin_2axes}). We can consider that the power
spectrum is damped by the factor $F(k,1-\nu^2)$ so that,
\begin{equation}  \label{eq:pk_two}
  P_{2\,{\rm axes}}^s(k)=\int_0^1d\nu\ \left\langle
    [\delta_g(k)+(1-\nu^2)\theta(k)]^2 F(k,1-\nu^2)\right\rangle
  \quad .
\end{equation}
For redshift-space distortions along three axes, whatever direction
${\bf k}/k$ we choose, we see redshift-space distortions as if we were
looking along the line of sight. So
\begin{equation} \label{eq:pk_three} 
  P_{3\,{\rm axes}}^s(k)=
  \left\langle [\delta_g(k)+\theta(k)]^2 F(k,1) \right\rangle 
  \quad .
\end{equation}

If we assume an Gaussian model for the SSVD damping, then
Eq.~(\ref{eq:pk_one}) can be solved using the $\gamma$ function,
\begin{equation} 
  \gamma(\alpha,x)\equiv\int_0^x\,dt\,e^{-t}t^{\alpha-1}
  \quad ,
\end{equation}
and the factors $1$, $2/3$ and $1/5$ in the expansion of
$P_{1\,{\rm axis}}^s(k)$ in Eq.~(\ref{eq:pk_sph_lin}) become
$\gamma(1/2,y^2)/(2y)$, $\gamma(3/2,y^2)/y^3$ and
$\gamma(5/2,y^2)/(2y^5)$ respectively. The original factors are
recovered in the limit $y\to0$. Alternative expressions can be derived
using error functions \citep{cole95,PeaDod96}.  Eq.~(\ref{eq:pk_two})
can be solved similarly using the imaginary error function.

If instead we assume an Exponential model for the small-scale velocity
dispersion, then the integral in Eq~(\ref{eq:pk_one}) can be
analytically solved using the $\tan^{-1}$ function \citep{cole95},
while the integral in Eq.~(\ref{eq:pk_two}) is trivial.

\section{Simulation results} \label{sec:sim_results}

\subsection{Fits to the spherically averaged power}

\begin{figure*}
\centering
\resizebox{0.32\textwidth}{!}{\includegraphics{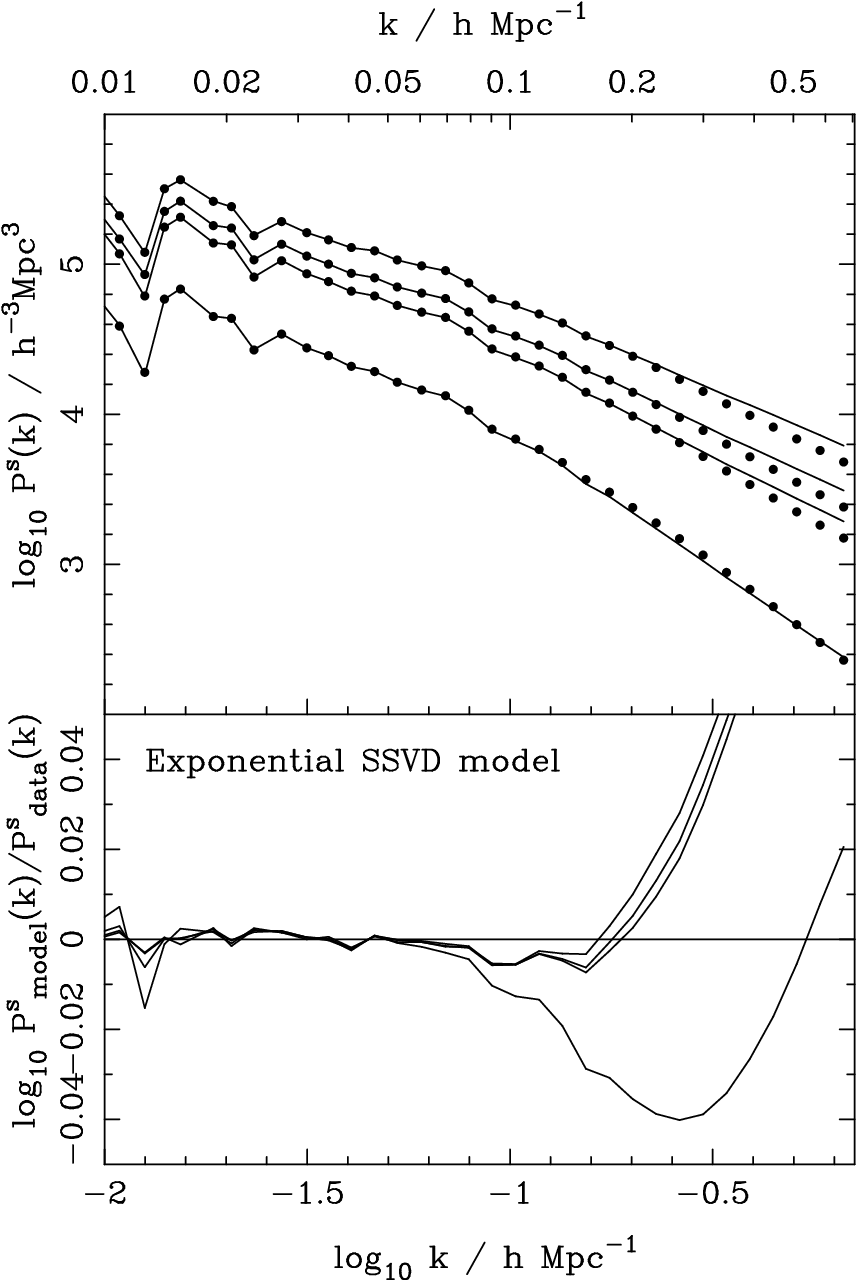}}
\hfill
\resizebox{0.32\textwidth}{!}{\includegraphics{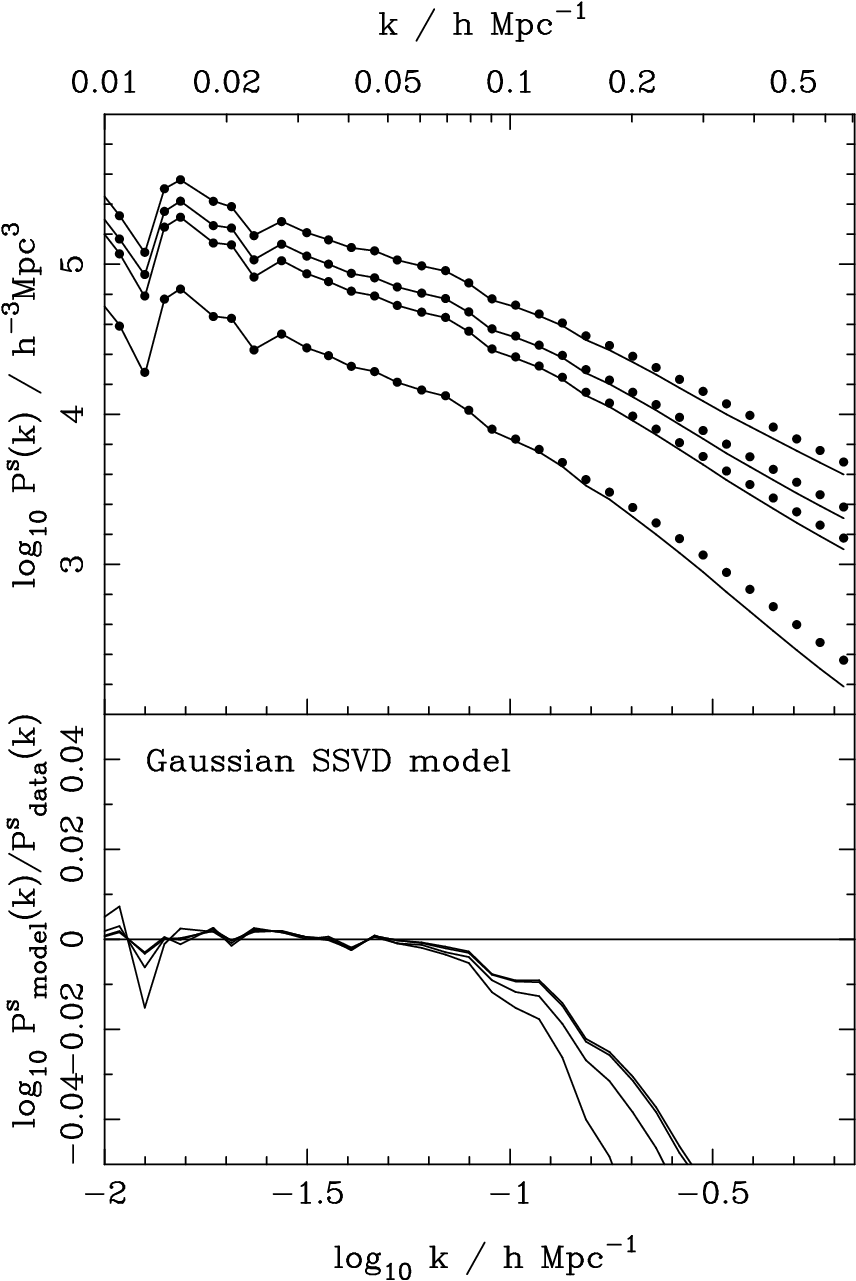}}
\hfill
\resizebox{0.32\textwidth}{!}{\includegraphics{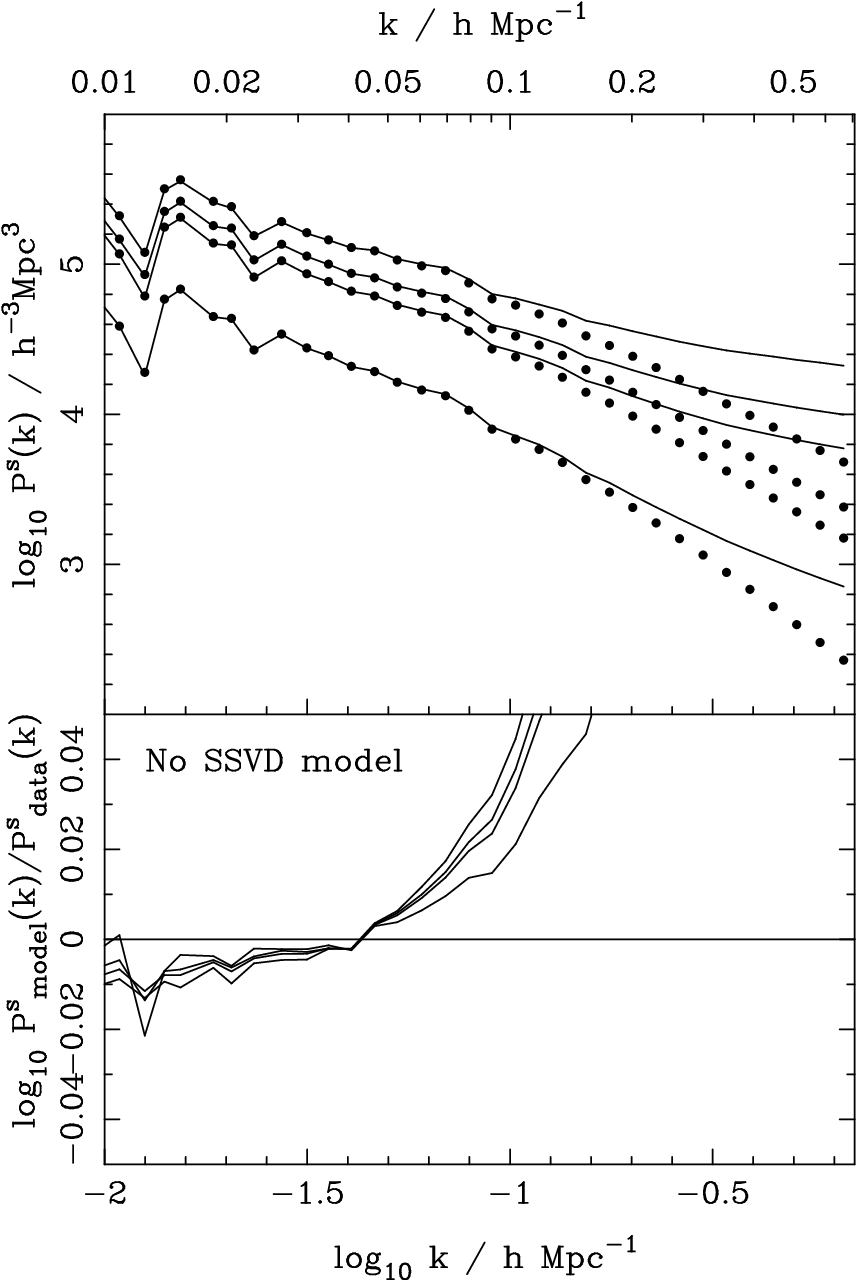}}
\caption{Recovered power spectra from our simulation, with
  redshift-space distortions along one axis of the simulation box
  (solid circles). Power spectra are shown for all mass (lower data),
  and for the three halo catalogues described in the text (upper data,
  with the more biased power spectra corresponding to larger mass
  thresholds). These data were fitted for $k<0.05\hompc$ using the
  model given by Eq.~(\ref{eq:pgs_comb_lin}). Ratios between model and
  data are shown in the lower panels, and panels from left to right
  are for different SSVD models. See Section~\ref{sec:sph_av_nl} for
  details.}
\label{fig:pk}
\end{figure*}

The first question we wish to address is ``How well does the linear
model with or without SSVD model (Eq.~\ref{eq:pgs_comb_lin}) recover
the monopole power $P_0^s=P_{1\,{\rm axis}}^s$?'' This comparison is shown
in Fig.~\ref{fig:pk}. The two
possible free parameters, $f/b$ and $\sigma$, were fitted on very
large scales $k<0.05\hompc$. The amplitude of the power spectrum on
these scales is close to the expected value for all data, following
linear theory (Section~\ref{sec:lin}) and given the simulation value
of $\beta$. Where no SSVD model is present, we see that the shape of
the model power spectrum does not match the data, even on these large
scales. Including either the Gaussian or Exponential SSVD model allows
a better match of the shape. The goodness-of-fit is approximately the
same for both of these models, which is not surprising as they have
the same asymptotic behaviour in the limit $y\to0$. High values of
$\sigma\sim 500$--$600 \kms$ are required to enable this match, which
fails for $k\simgt0.1\hompc$. While the SSVD model was introduced to
correct the high-$k$ behaviour of redshift-space power spectra, it is
clearly having an effect on scales $0.05<k<0.1\hompc$, where it is
attempting to correct both the quasi-linear redshift-space effects and
the deterministic linear bias assumption.

\begin{figure*}
\centering
\resizebox{0.32\textwidth}{!}{\includegraphics{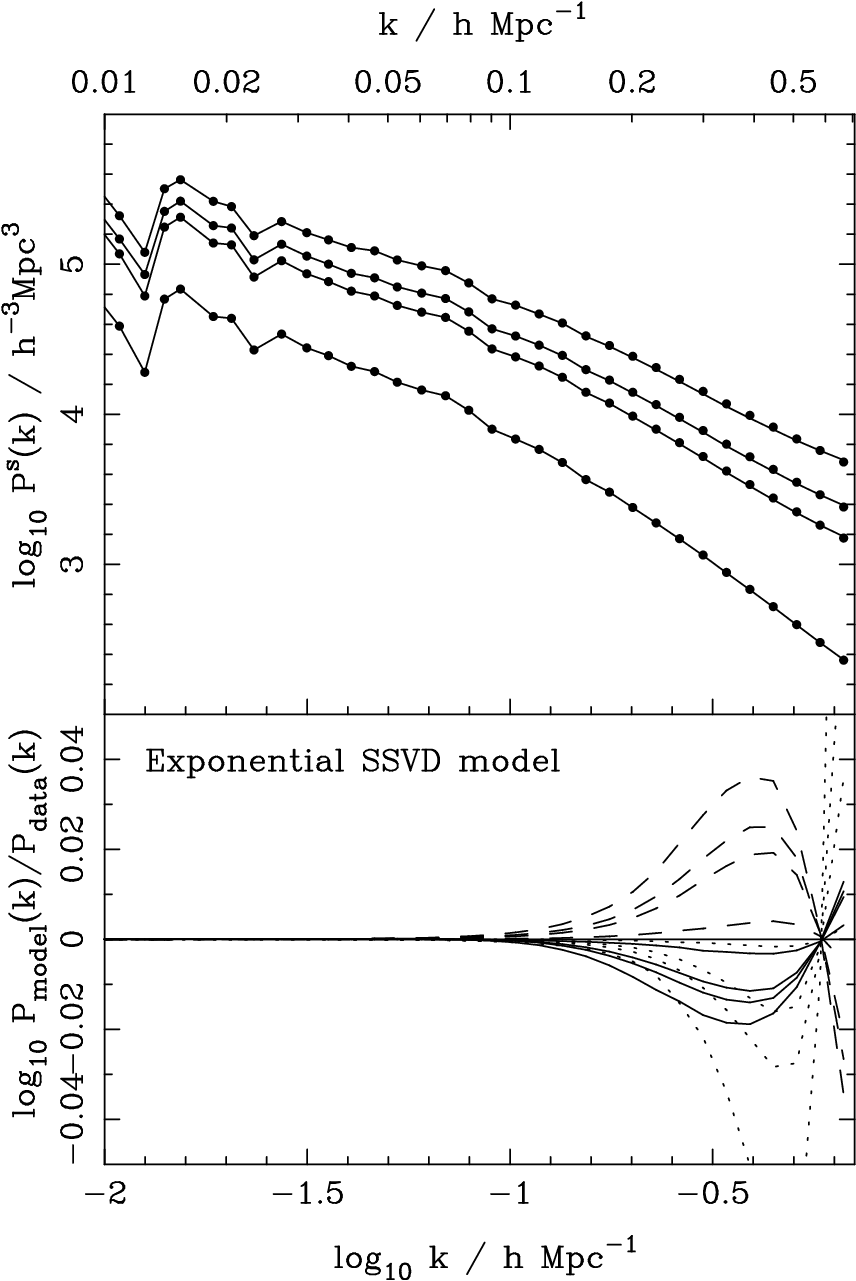}}
\hfill
\resizebox{0.32\textwidth}{!}{\includegraphics{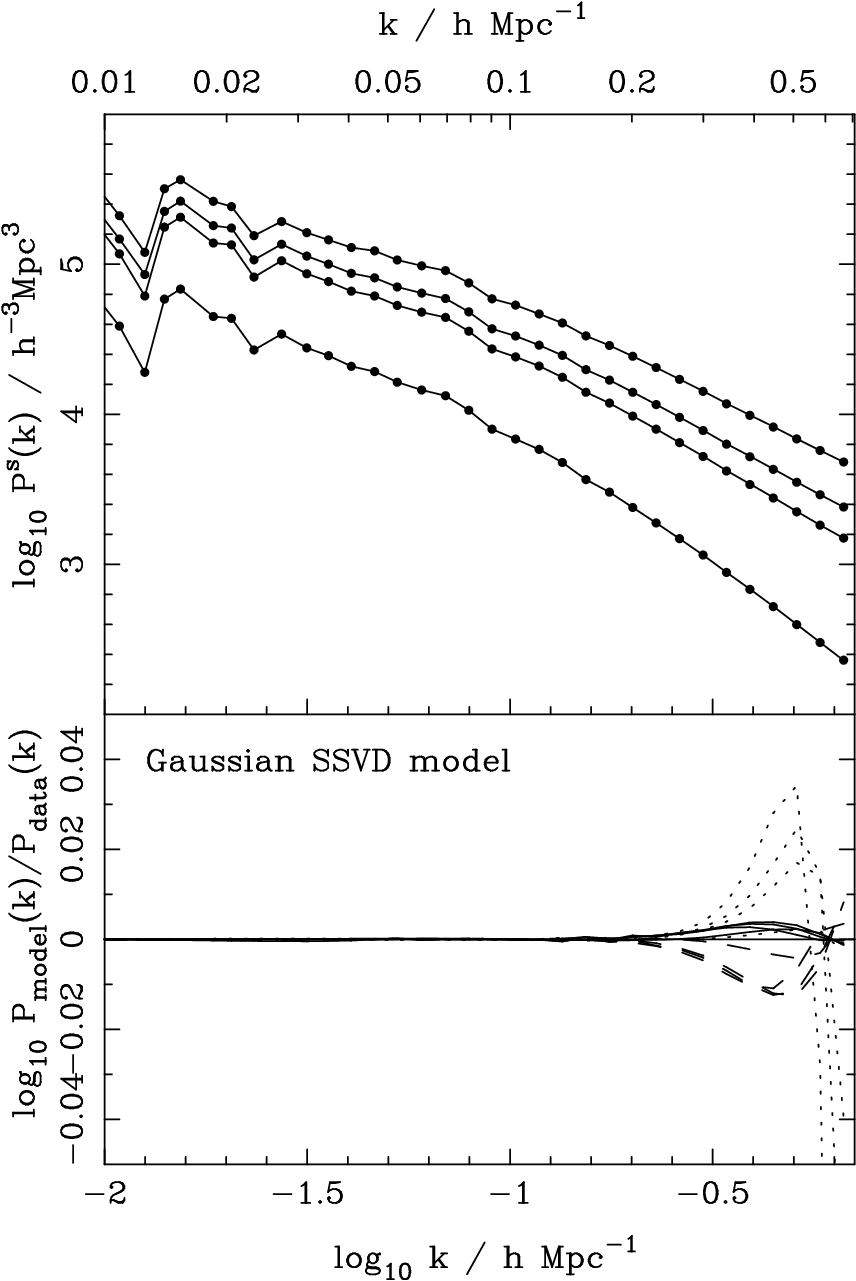}}
\hfill
\resizebox{0.32\textwidth}{!}{\includegraphics{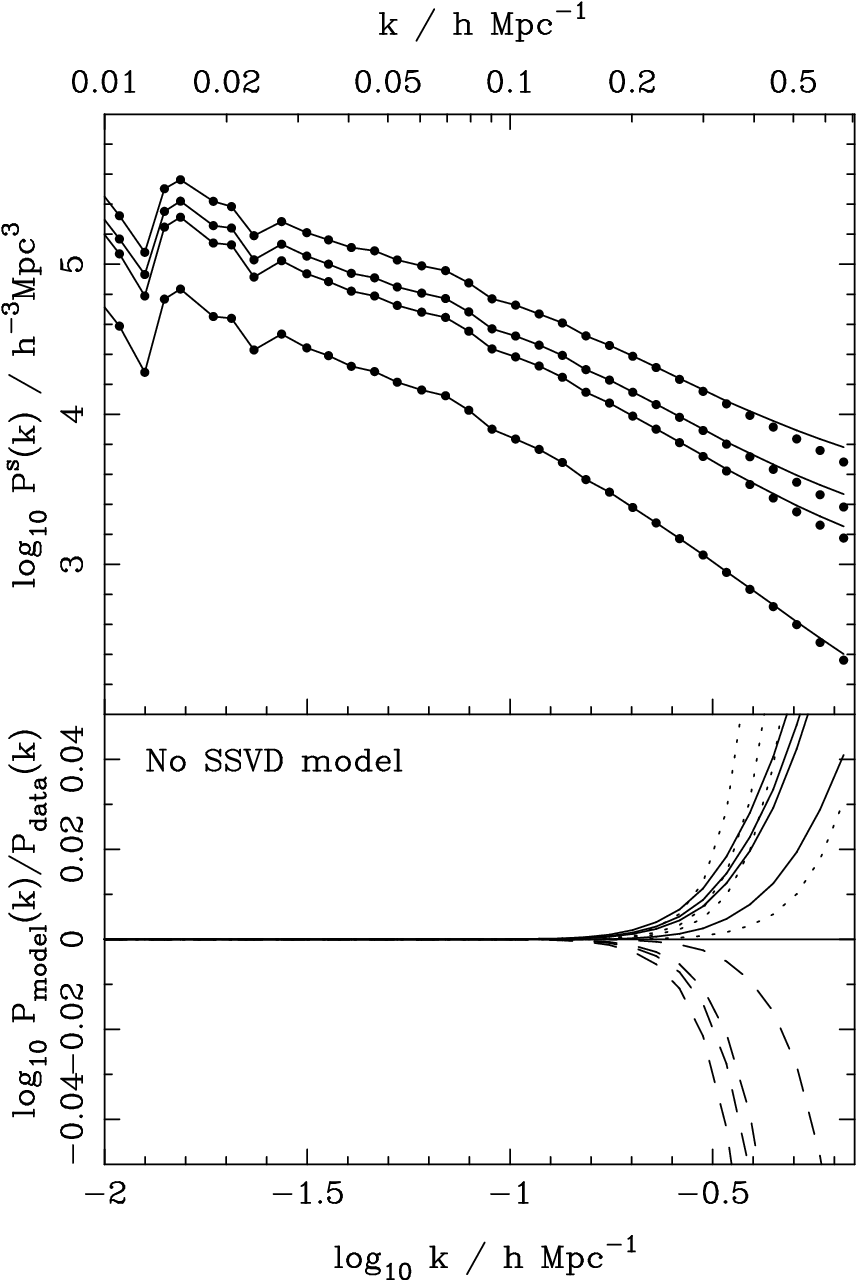}}
\caption{$P_{1\,{\rm axis}}^s(k)$ calculated from the simulation are
  shown by the solid circles - these are the same data in the upper
  panels of Fig.~\ref{fig:pk}. The lowest amplitude power spectrum
  corresponds to the mass, while the other three are from halo
  catalogues, as described in the text, where the amplitude is an
  increasing function of mass limit. For comparison we also plot the
  model given by Eqns.~\ref{eq:pk_one}, where the free parameters
  fitted are $P_{gg}$, $P_{g\theta}$, $P_{\theta\theta}$, and $\sigma$
  where a SSVD model is included (solid lines). To determine these
  models, we have fitted $P_{1\,{\rm axis}}^s$, $P_{2\,{\rm axes}}^s$,
  and $P_{3\,{\rm axes}}^s$ for $k<0.7\hompc$, where the data were
  equally weighted in $\log k$. In order to highlight differences, we
  plot the model for $P_{1\,{\rm axis}}^s$ divided by the data in the
  lower panels (solid lines).  Similar lines are also plotted for
  $P_{2\,{\rm axes}}^s$ (dashed), and $P_{3\,{\rm axes}}^s$ (dotted).}
\label{fig:pk_axes}
\end{figure*}

The second question that we wish to answer is ``How well does the
model of Eq.~(\ref{eq:pgs_combined}) where $P_{gg}$, $P_{g\theta}$ \&
$P_{\theta\theta}$ are arbitrary, with or without a SSVD model, fit
the simulation results?''. The power spectra, $P_{1\,{\rm axis}}^s$
are shown by the solid circles in the upper panels of
Fig.~\ref{fig:pk_axes}. On large-scales, $k<0.1\hompc$, the data are
well fitted by the model of Eq.~(\ref{eq:pgs_combined}) without the
inclusion of a SSVD component. As we move to scales $k>0.1\hompc$, we
need to include a SSVD model in order to match the differences between
$P_{1\,{\rm axis}}^s$, $P_{2\,{\rm axes}}^s$, and $P_{3\,{\rm
    axes}}^s$, which result from the $\mu$ dependence of
$P_g^s(k,\mu)$. Ratios between model and data for $P_{1\,{\rm
    axis}}^s$, $P_{2\,{\rm axes}}^s$, and $P_{3\,{\rm axes}}^s$ are
shown in the lower panels of this plot. Although the SSVD models
significantly help with the fits to the observed power spectra, it is
clear that the shapes cannot be matched and that a compromise is being
reached in the fits, where the model crosses the data at
$k\sim0.6\hompc$. The best-fit values of $\sigma$ vary with the range
of scales fitted confirming this picture. On these quasi-linear
scales, the Gaussian SSVD model allows a slightly better fit to the
data, although it is not clear that there is a physical motivation
behind this. The Exponential model is a better fit on smaller scales,
which is physically motivated \citep{PeaDod96,Whi01}.

\subsection{Extracting density and velocity power spectra}

\begin{figure*}
\centering
\resizebox{0.32\textwidth}{!}{\includegraphics{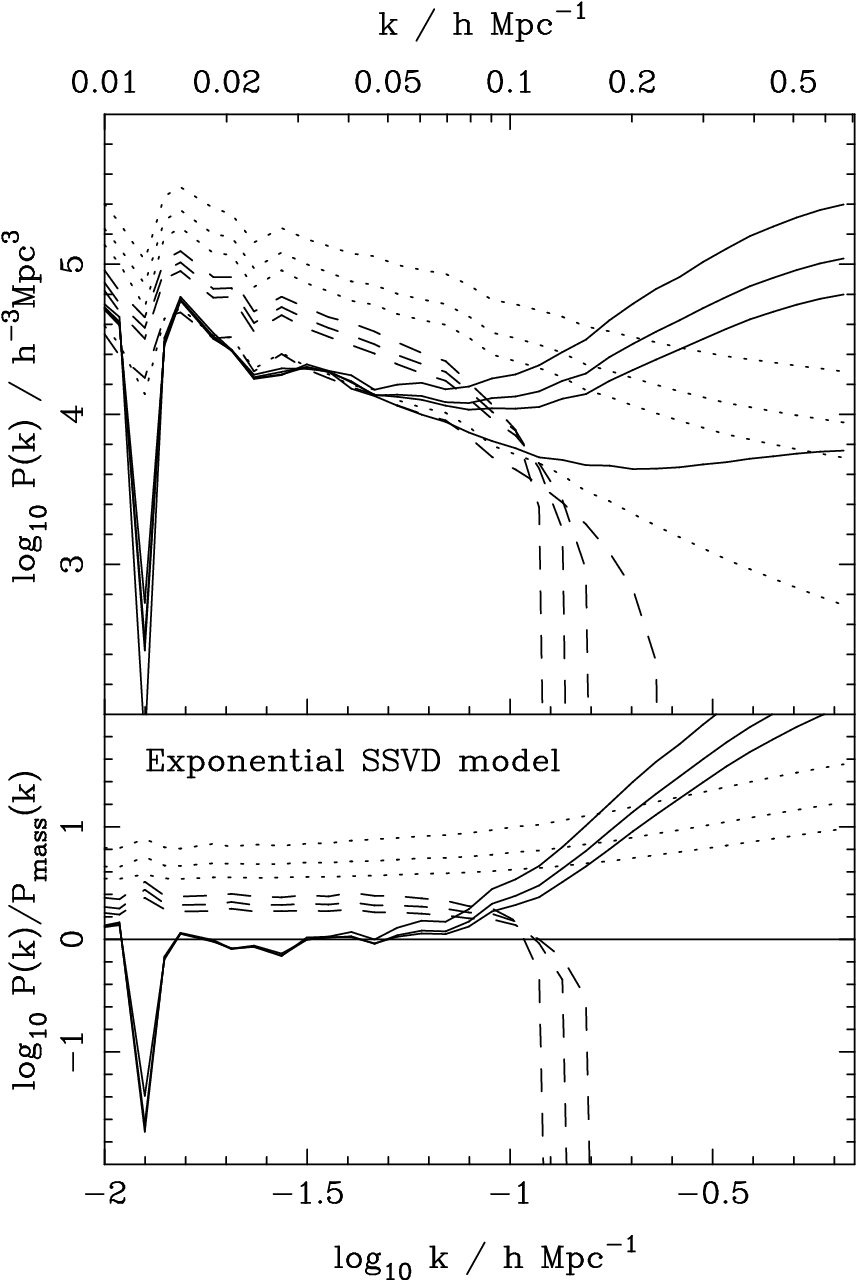}}\hfill
\resizebox{0.32\textwidth}{!}{\includegraphics{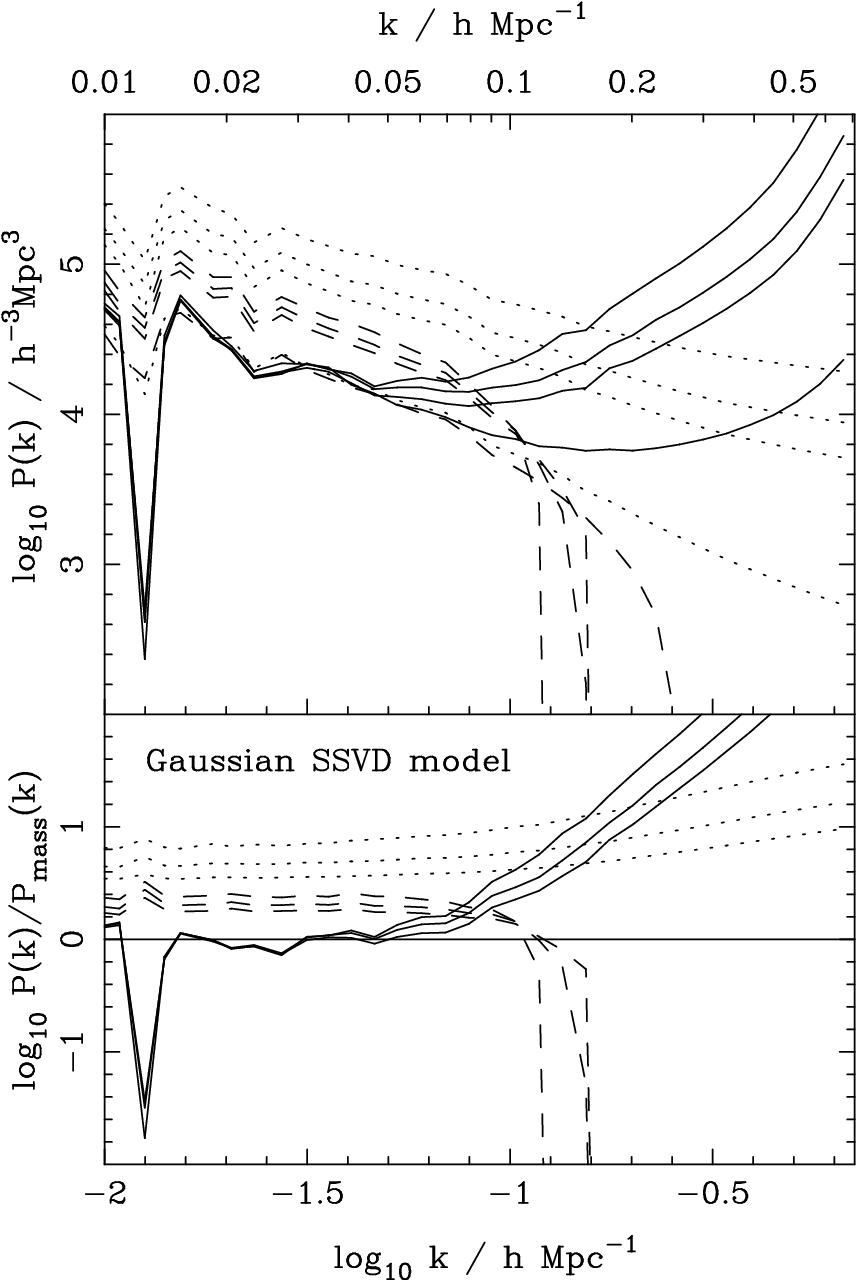}}
\hfill
\resizebox{0.32\textwidth}{!}{\includegraphics{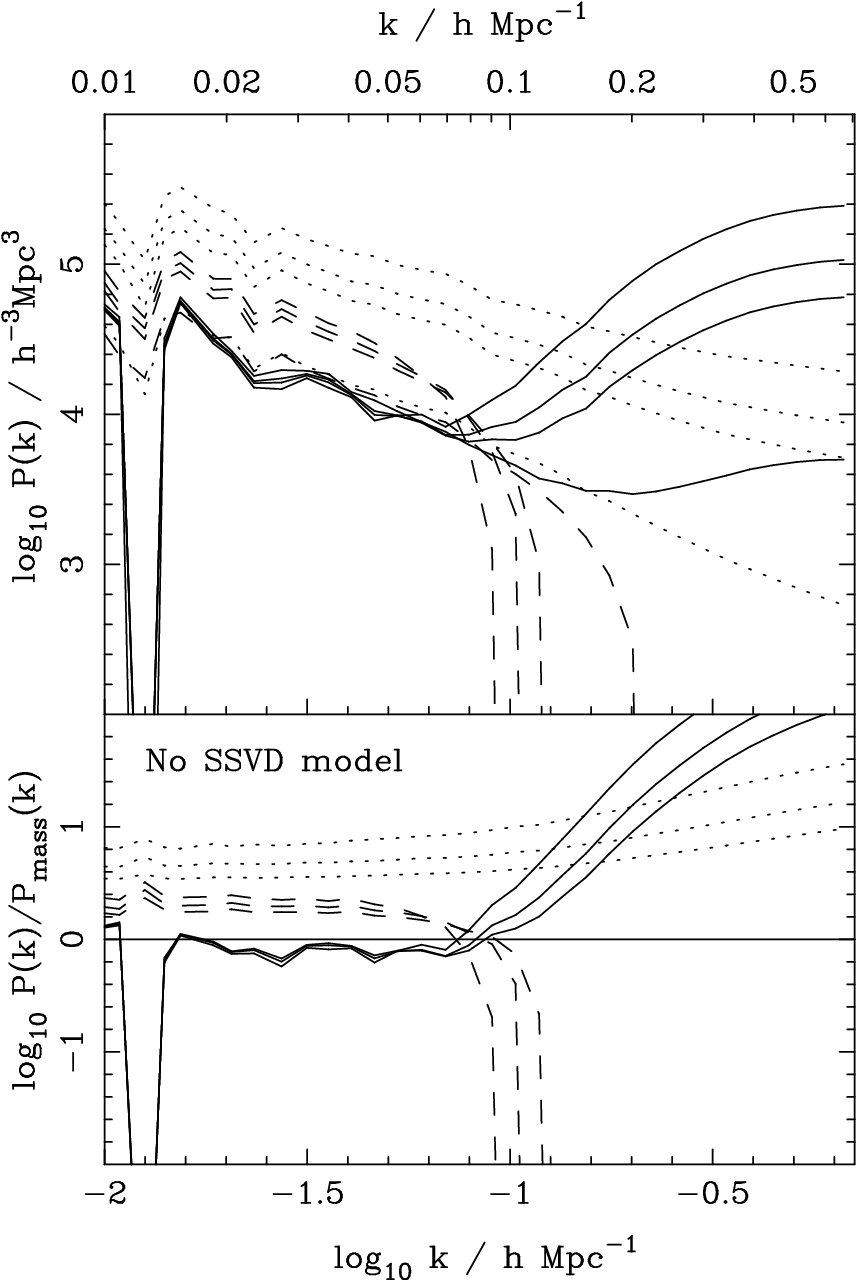}}
\caption{Upper panels: $P_{gg}$ (dotted lines), $P_{g\theta}/f$
  (dashed lines), $P_{\theta\theta}/f^2$ (solid lines) recovered by
  fitting $P_{0\,{\rm axes}}^s$, $P_{1\,{\rm axis}}^s$, $P_{2\,{\rm
      axes}}^s$, and $P_{3\,{\rm axes}}^s$ for $k<0.7\hompc$. Four
  power spectra are plotted, corresponding to the mass and the halo
  catalogues. The amplitude of $P_{gg}$ and $P_{g\theta}$ is lowest
  for the mass, and is an increasing function of the halo mass
  limit. For $P_{\theta\theta}$, the deviation from the shape of
  $P_{gg}$ is weakest for the mass, and is an increasing function of
  halo mass limit. The lower panels show these power spectra divided
  by $P_{gg}$ for the mass to highlight deviations between the shapes
  of the power spectra.}
\label{fig:pk_split}
\end{figure*}

The third question that we want to use the simulations to answer is
``How are the power spectra $P_{gg}$, $P_{g\theta}$ \&
$P_{\theta\theta}$ related to each other and to the matter power
spectrum?''. To fit for $P_{gg}$, $P_{g\theta}$ \& $P_{\theta\theta}$
we use spherically averaged power spectra calculated assuming that we
have redshift-space distortions along 0, 1, 2 or 3 axes. Without
including a SSVD model, we could use any three of $P_{0\,{\rm
    axes}}^s$, $P_{1\,{\rm axis}}^s$, $P_{2\,{\rm axes}}^s$, and
$P_{3\,{\rm axes}}^s$ to solve Eq.~(\ref{eq:pk_sph_lin}) for $P_{gg}$,
$P_{g\theta}$ and $P_{\theta\theta}$. In order to simplify the fit, we
assumed $P_{gg}=P_{0\,{\rm axes}}^s$, and used $P_{1\,{\rm axis}}^s$
and $P_{2\,{\rm axes}}^s$ to solve for $P_{g\theta}$ \&
$P_{\theta\theta}$. For fits including a SSVD model, we used all four
power spectra to solve for $\sigma$, $P_{gg}$, $P_{g\theta}$ and
$P_{\theta\theta}$.

The power spectra, $P_{gg}$, $P_{g\theta}/f$, $P_{\theta\theta}/f^2$,
recovered from the fits to $P_{i\,{\rm axis}}^s$ are plotted in the
upper panels in Fig.~\ref{fig:pk_split}. The power spectra divided
by $P_{\rm mass}$ are shown in the lower panels to highlight
variations in shape of these different power spectra. The residuals
between model and simulation data for $P_{i\,{\rm axis}}^s$ were shown
in Fig.~\ref{fig:pk_axes}. On the largest scales $k<0.05\hompc$, the
data are consistent with the hypothesis that $P_{\theta\theta}\simeq
f^2P_{\rm mass}$, and $P_{g\theta}^2\simeq f^2P_{gg}P_{\rm
  mass}$. Where we do not include a SSVD model, there is weak evidence
that the amplitude of $P_{\theta\theta}/f^2$ is lower than the
amplitude of the matter power spectrum on scales $k<0.05\hompc$,
consistent with a 10\% velocity bias, although it should be noted that
the data are noisy here.

It is clear that the shapes of the velocity and overdensity power spectra
do not match on scales $k>0.05\hompc$, even when either Gaussian or Exponential
SSVD models are included.  Deviations between the shapes of these power
spectra were also noted in Figure 6 of \citet{Sco04}, and were physically
motivated by the arguments in Section~\ref{sec:qlin}.
The turn-off from linear behaviour is stronger for $P_{\theta\theta}$ than
$P_{gg}$ for any of the catalogues, which is consistent with the quasi-linear
particle velocities leading the displacements, which is an integral over the
velocities.

\begin{figure}
\centering
\resizebox{0.9\columnwidth}{!}{\includegraphics{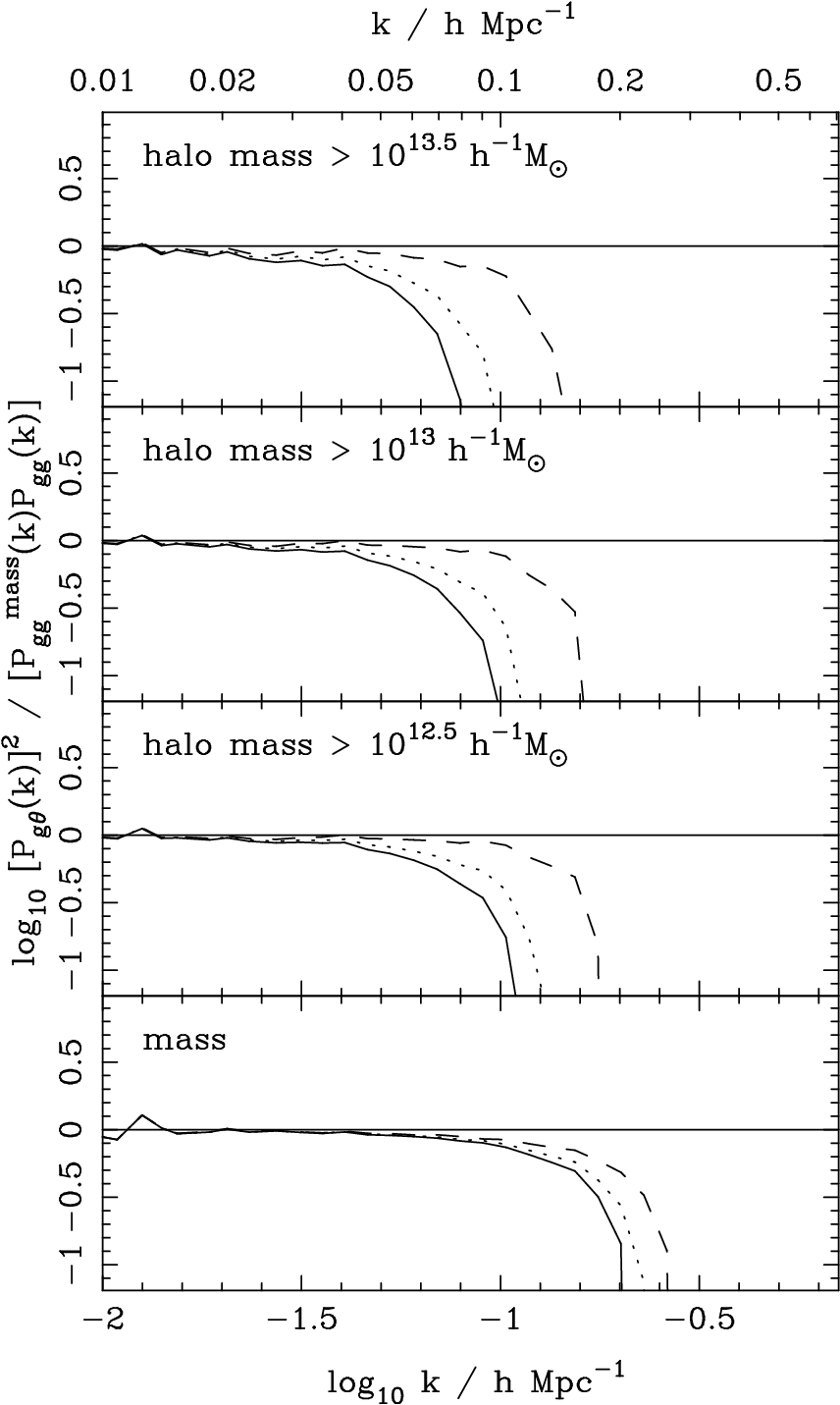}}
\caption{For $P_{g\theta}$, we can remove a deterministic linear
  galaxy density bias dependence by dividing by $\sqrt{P_{gg}}$. If
  the galaxy density bias is well described by a local linear model
  and there is no velocity bias, then we expect
  $P^2_{g\theta}/P_{gg}=P_{\rm mass}$, so we divide $P_{g\theta}$ by a
  further factor of $\sqrt{P_{\rm mass}}$ to highlight deviations from
  this model. The solid line was calculated from a fit with no
  small-scale dispersion correction, the dotted line assumed an
  Exponential model, and the dashed line a Gaussian model, both with
  single scale-independent variance.}
\label{fig:pk_Pgt}
\end{figure}

As we discuss in Section~\ref{sec:estimators}, most of the
cosmological signal from observations comes from the cross correlation
between overdensity and velocity fields. So, the forth question that
we wish to use simulations to answer is ``Is
$P_{g\theta}^2/(P_{gg}P_{\rm mass})=1$ a valid model on large
scales?''. To answer this, we plot the power spectrum combination
$P_{g\theta}^2/(P_{gg}P_{\rm mass})$ in Fig.~\ref{fig:pk_Pgt}.  On
large scales, this statistic $\to1$, validating our
hypothesis. Comparison with Fig.~\ref{fig:pk_split} suggests that it
is less biased than $P_{\theta\theta}$ as a tracer of
$f^2\sigma_8^2({\rm mass})$ on these scales, as the deviations from
$P_{g\theta}^2/(P_{gg}P_{\rm mass})=1$ are less than 10\%. The
large-scale ``plateau'' can be extended by assuming a Gaussian model
for the small-scale velocity dispersion, provided the fit is not
extended to large $k$. For the plot, we only fitted to $k<0.4\hompc$.
The recovered values of $\sigma$ for Gaussian model of
scale-independent velocity dispersion correction are $310\,{\rm
  kms}^{-1}$, $360\,{\rm kms}^{-1}$, $390\,{\rm kms}^{-1}$, $470\,{\rm
  kms}^{-1}$, for the mass, and the three halo catalogues. For the
Exponential model, the corresponding numbers are $280\,{\rm
  kms}^{-1}$, $300\,{\rm kms}^{-1}$, $340\,{\rm kms}^{-1}$, $410\,{\rm
  kms}^{-1}$. If we extend the fit to smaller scales, then $\sigma$
increases for both models and the results from Exponential and
Gaussian models become more similar. In addition, the plateau is
reduced in size.

The sharp decrease in $P_{g\theta}$ seen at $k\sim0.1\hompc$ is
consistent with a model that includes both the effect of FOGs and the
infall model. The inflection marks the point where the highest
galaxies velocities change from being associated with the largest
scales (due to infall), to instead be linked to the smallest scales
(due to FOG). The quasi-linear nature of infall on small scales will
also contribute to this transition. The scale at which the
inflection occurs is affected by the SSVD model included. The Gaussian
SSVD model acts as a slightly more accurate fitting function compared
with the exponential model, extending the plateau of where the
overdensity-velocity cross power spectrum matches the shape of
$P_{gg}$ to $k<0.1\hompc$. This might reflect the contribution of
quasi-linear effects to the FOG, where the Exponential model was
physically motivated. We now investigate the SSVD model further.

\subsection{Investigating SSVD fits}

\begin{figure}
\centering
\resizebox{0.9\columnwidth}{!}{\includegraphics{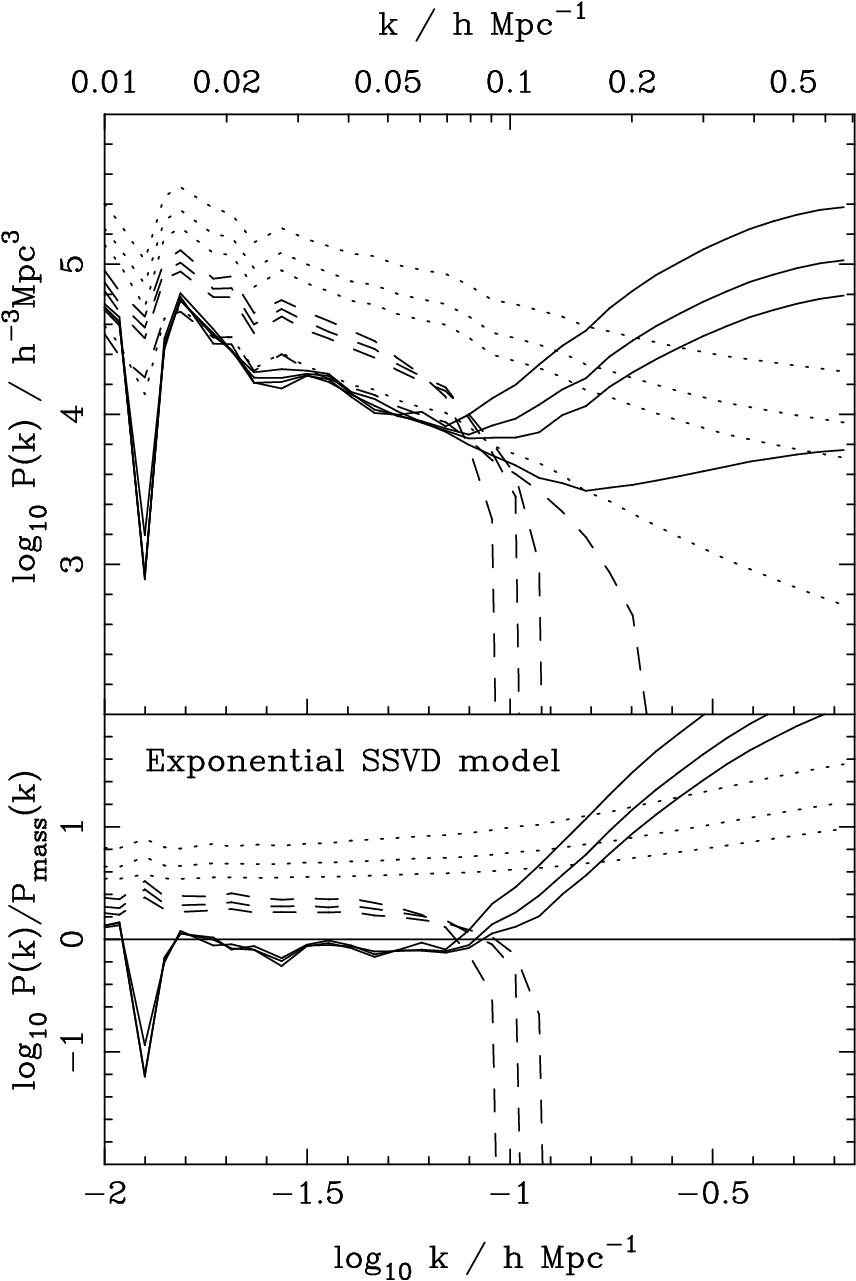}}
\caption{As Fig.~\ref{fig:pk_split}, but now allowing the variance of
  the assumed SSVD model, $\sigma$ to vary with scale. The resulting
  models of $P_{1\,{\rm axis}}^s$, $P_{2\,{\rm axes}}^s$, and
  $P_{3\,{\rm axes}}^s$ are an almost perfect fit to the data. This
  should be expected as we are fitting 4 data points with a model with
  4 parameters at each $k$-value.}
\label{fig:pk_split_vs}
\end{figure}

The fifth question that we wish to address using simulations is ``How
well do the Gaussian and Exponential SSVD models in
Eq.~(\ref{eq:pgs_combined}) extend the fit to small scales?''. This is
clearly a difficult question to answer, but we have tried to quantify
the effect, and determine the scales where the SSVD model is
important, by allowing the free parameter, $\sigma$ in the Gaussian
and Exponential SSVD models to vary as a function of scale. The power
spectra $P_{gg}$, $P_{g\theta}/f$ \& $P_{\theta\theta}/f^2$,
calculated assuming the Exponential SSVD model are shown in
Fig.~\ref{fig:pk_split_vs}, which has the same format as
Fig.~\ref{fig:pk_split}. Allowing $\sigma$ to vary allows a virtually
exact fit to $P_{1\,{\rm axis}}^s$, $P_{2\,{\rm axes}}^s$, and
$P_{3\,{\rm axes}}^s$. For $k<0.1\hompc$, Fig.~\ref{fig:pk_axes} shows
that no SSVD model is required to fit the observed power spectra, and
there is no constraint on $\sigma$. For $k>0.1\hompc$ the best fit
value of $\sigma$ rises from $0$ at $k=0.1\hompc$ to $295\kms$ at
$k=1\hompc$ for the matter, and to $330\kms$, $390\kms$, and $470\kms$
for the halo catalogues with mass $>10^{12.5}$, $10^{13}$ \&
$10^{13.5}\,h^{-1}\msun$ respectively. A similar trend is seen for the
Gaussian model, although there is more noise, perhaps indicating that
the parameters $P_{gg}$, $P_{g\theta}$, \& $P_{\theta\theta}$ and
$\sigma$ fitted for each $k$ are more degenerate in this model. This
would also explain why the Gaussian model with a single value of
$\sigma$ for all scales can provide a slightly better fit to the data
shown in Fig.~\ref{fig:pk_axes}: the solution can move along the
Likelihood degeneracy.

\subsection{Testing our new estimator} \label{sec:sim_test_estimator}

\begin{figure}
\centering
\resizebox{0.9\columnwidth}{!}{\includegraphics{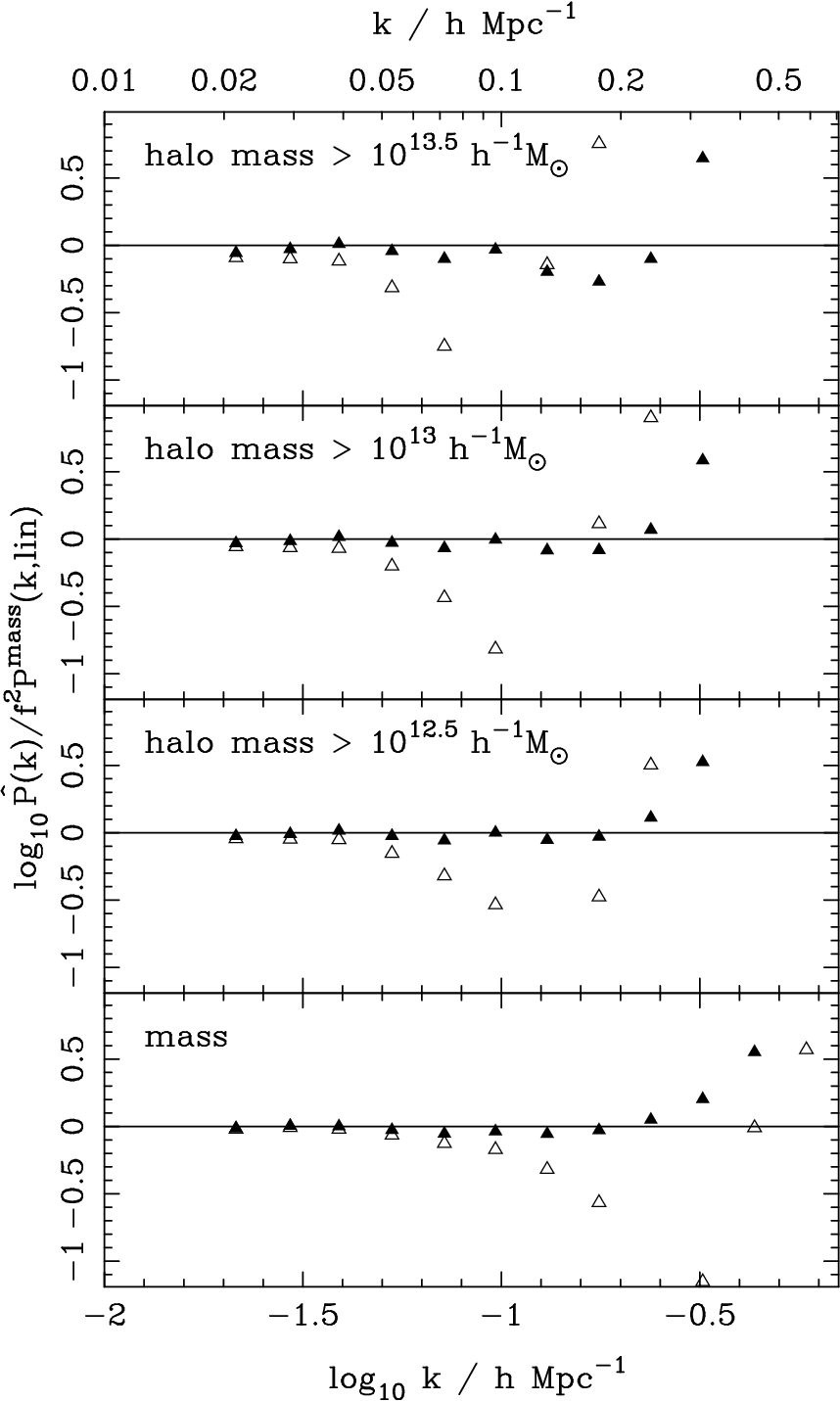}}
\caption{The power spectrum given by Eq.~(\ref{eq:bsig8}) divided by
  the real-space linear power spectrum multiplied by $f^2$ (open
  triangles). On large scales $k<0.05\hompc$ this tends to 1 as
  expected. For comparison, the solid triangles give the value of the
  estimator given by Eq.~(\ref{eq:pk_leg_gauss}), where we include a
  Gaussian component equivalent to a SSVD model, with variance fitted
  to $P_{1\,{\rm axis}}^s$, $P_{2\,{\rm axes}}^s$, and $P_{3\,{\rm axes}}^s$
  (see text for details). This demonstrates that a Gaussian SSVD model
  on scales $k<0.4\hompc$ can be treated as a fitting formulae for the
  quasi-linear turn-off from the expected linear behaviour.}
\label{fig:pk_leg}
\end{figure}

The final question that we wish to answer from simulations is ``How
well do the power spectra constructed from our new estimators given in
Eqns.~(\ref{eq:pk_from_legendre}) \& (\ref{eq:pk_leg_gauss}), recover
$f^2P_{\rm mass}(k)$?''. The power spectra are compared in
Fig.~\ref{fig:pk_leg}. Even with only $P_0^s$ and $P_2^s$, we see that
we can recover an unbiased estimate of $f^2P_{\rm mass}(k)$ on
large-scales. The inclusion of a Gaussian SSVD model is able to
correctly model the quasi-linear behaviour of this estimator for
$k<0.2\hompc$ for the mass, a limit that decreases to $k<0.1\hompc$
for the most massive halo catalogue. The value of $\sigma$ can
therefore be treated as a free ``nuisance'' parameter, and could be
fitted to the estimator. In order to demonstrate consistency we
actually plot the estimator as in Eq.~(\ref{eq:pk_leg_gauss}), with
$\sigma$ fitted to $P_{1\,{\rm axis}}^s$, $P_{2\,{\rm axes}}^s$, and
$P_{3\,{\rm axes}}^s$ for $k<0.4\hompc$. In practice, for a given
linear matter power spectrum, one could calculate the estimator for a
series of values of $\sigma$ and marginalise over this parameter when
fitting to the model.

Comparing the estimator of Eq.~(\ref{eq:pk_leg_gauss}) to the
non-linear matter power spectrum would result in a better fit to the
data, because the behaviour of $\hat{P}$ and the non-linear matter
power spectrum are similar. We consider that this would be artificial,
because we do not expect such a match from the theory presented in
Section~\ref{sec:zspace_theory}. With the Gaussian damping term, our
new estimator is clearly a complicated function of $P_0^s$ and
$P_2^s$, and it would be difficult to accurately propagate standard
analytic error estimates for power spectrum modes to this
function. Given the precision to which future surveys would be able to
measure this statistic, it would be better to base error analyses on
the results from mock catalogues.

\section{Discussion}

Redshift-space distortions encode key information about the build-up
of cosmological structure. In the linear regime, the theory behind the
density enhancement was developed over 20 years ago \citep{Kai87}.
Following this development, most observational studies
have focused on measuring $\beta$. One method to do this is to measure
the quadrupole-to-monopole ratio, which depends on $\beta$ in a known
way (Eq.~\ref{eq:P2overP0}). By reviewing the theory of redshift-space
distortions in Section~\ref{sec:zspace_theory}, we argued that we
should be able to obtain a bias independent cosmological constraint on
$f\sigma_8({\rm mass})$ from the large-scale redshift-space
distortions.
This statistic, proportional to $dD/d\ln a$ with a proportionality constant
depending on the amplitude of fluctuations at early times, provides an
excellent test of DE models \citep[e.g.][]{song08b}.

In Section~\ref{sec:qlin} \&~\ref{sec:nlin}, we set out to extend the
theory into the quasi- and non-linear regimes, to investigate the
smallest scales on which we can easily recover cosmological
information using this physical process. In the linear regime, the
power spectrum can be decomposed into terms dependent on $\mu^0$,
$\mu^2$ and $\mu^4$. As we push to quasi-linear scales we should
expect this behaviour to become more complicated with extra $\mu^6$
and higher order terms becoming important. We should also expect the
simple relations $bP_{g\theta}=fP_{gg}$ and
$b^2P_{\theta\theta}=f^2P_{gg}$ to break down. Previous work has
introduced a ``streaming'' model for non-linear
redshift-space distortions, including a damping term spliced together
with linear theory. In the limit $k\to0$, the standard Gaussian or
Exponential forms given in Eqns.~(\ref{eq:F_exp})
\&~(\ref{eq:F_gauss}) for this damping function both reduce to
providing an extra $\mu^6$ term with amplitude dependent on
$\sigma$. We might therefore expect these terms can also be used to
match the expected quasi-linear $\mu$ dependence of the power
spectrum. In this case, the same function might not match all
quasi-linear and non-linear scales.

Using a large numerical simulation we have set out to test this
theory. To facilitate this, we have introduced a novel approach to the
analysis of the redshift-space power spectrum measured from
simulations by allowing redshift-space distortions to be included
along multiple axes. Power spectra calculated from density fields
including redshift-space distortions along multiple axes act as
moments of $P_g^s(k,\mu)$, allowing us to decompose into the expected
$\mu^0$, $\mu^2$ and $\mu^4$ components. On large scales we see no
evidence for strong velocity bias, and the velocity-velocity power
spectrum has an amplitude $f^2$ times that of the mass to within 10\%
(right panel of Fig.~\ref{fig:pk_split}). The density-velocity power
spectrum, from which most information is obtained from surveys is less
affected by this velocity bias. Fig.~\ref{fig:pk} shows that the
relations $bP_{g\theta}=fP_{gg}$ and $b^2P_{\theta\theta}=f^2P_{gg}$
break down on scales $k>0.1\hompc$. The right panel of
Fig.~\ref{fig:pk_axes} shows that $\mu^6$ and higher order terms need
to be included for $k>0.2\hompc$. The Exponential and Gaussian models
enable a better fit for $0.2<k<0.4\hompc$, but neither is perfect on
all of these scales, although the Gaussian model does slightly better
than the Exponential model in the cases we illustrate. Alternative models
which reduce to $\mu^6$ behaviour in the limit $k\to0$ might perform better,
but this is not necessary for our purposes: we have shown that the
Gaussian model extends the range of the fit.

Based on this analysis, we have devised a new estimator of the matter
power spectrum $\hat{P}$ using the monopole and quadrupole power
spectra. This matches the large-scale shape of the matter power
spectrum with amplitude multiplied by $f^2$, so we can measure
$f\sigma_8({\rm mass})$. Including the Gaussian damping model allows
us to extend the match with the simulation data for
$k\simlt0.2\hompc$. The match between the shape of the expected linear
matter power spectrum and $\hat{P}$ should allow a test of the scales
over which we can measure $f\sigma_8({\rm mass})$.  Rather than use
such an estimator, it would also be possible to perform a maximum
likelihood fit of Eq.~(\ref{eq:pgs_comb_lin}) to the measured
redshift-space power spectrum, with $f\sigma_8({\rm mass})$ and
$b\sigma_8({\rm mass})$ as the free parameters.

Our analysis has assumed the distant observer limit, a standard
approximation used in many analyses. It would be straightforward to
extend this work to methods that incorporate a proper radial--angular
split allowing for standard survey geometries. An additional extension
would be to perform such an analysis in configuration space: again, it
would also be possible the extend the ideas behind our estimator to
analyse the correlation function.

\section*{Acknowledgments}

We thank the referee Andrew J.S. Hamilton for carefully reading our
manuscript and his insightful and concrete suggestions for
improvements. We thank Jeremy Tinker and Luigi Guzzo, for helpful
comments on an earlier draft of this paper. WJP is supported by STFC,
the Leverhulme Trust and the European Research Council. WJP thanks
Lawrence Berkeley National Laboratory for their hospitality during a
visit in May 2008, when this work was initiated. MW is supported by
NASA. The simulations used in this paper were analyzed at the National
Energy Research Scientific Computing Center.

\label{lastpage}
\end{document}